\documentclass{article}

\usepackage{PRIMEarxiv}
\usepackage[numbers]{natbib}
\usepackage{authblk}
\usepackage{enumitem}
\usepackage[utf8]{inputenc} 
\usepackage[T1]{fontenc}    
\usepackage{hyperref}       
\usepackage{url}            
\usepackage{booktabs}       
\usepackage{multirow}
\usepackage{amsfonts}       
\usepackage{nicefrac}       
\usepackage{microtype}      
\usepackage{amsmath}
\usepackage{amssymb}
\usepackage{fancyhdr}       
\usepackage{graphicx}       
\usepackage{subcaption}
\usepackage{tabularx}
\usepackage{array}
\usepackage{algorithm}
\usepackage{algpseudocode}
\usepackage{pifont}
\graphicspath{{media/}}     

\title{DDIM-Driven Coverless Steganography Scheme with Real Key}

\author{
  Mingyu Yu, Haonan Miao, Zhengping Jin, Sujuan Qin\thanks{Corresponding author}\\
  State Key Laboratory of Networking and Switching Technology\\
  Beijing University of Posts and Telecommunications, Beijing\\
  \texttt{\{yumingyu, hnmiao0314, zhpjin, qsujuan\}@bupt.edu.cn}\\
}

\begin{document}
\maketitle
\nocite{*}

\begin{abstract}
With the advancement of information hiding techniques, generation-based coverless steganography has emerged as an alternative to traditional methods, leveraging generative models to transform secret information into stego-objects rather than embedding it within the redundancy of the cover.
However, existing generation-based approaches require pseudo-keys that must be shared between communication parties, leading to high overhead of frequent key exchanges and security risks due to their tight coupling with the secret information.
This paper proposes a DDIM-driven coverless steganography scheme that utilizes a real-key mechanism, improving the key management.
By integrating reversible data hiding (RDH) and chaotic encryption into generation model, the proposed method eliminates excessive key exchanges and reduces the correlation between the key and the secret information.
Furthermore, it requires only a single key negotiation for multiple communication, which lowers overhead.
Experimental results demonstrate that the proposed scheme resists substitution attacks, enhancing the security of covert communication.
\end{abstract}


\section{Introduction}
As secure communication becomes increasingly vital in the digital era, steganography has gained significant attention for its ability to conceal sensitive information within various media.
Typical steganographic methods embed secret information into a cover by modifying its statistical or perceptual attributes, which can compromise security and flexibility.
In contrast, coverless steganography eliminates such modifications by leveraging intrinsic data properties to generate or select cover images, enhancing both security and adaptability.

The emergence of advanced generative models has further transformed coverless steganography.
Generative Adversarial Networks (GANs) \cite{goodfellow2014gan}, known for their ability to synthesize high-quality images, were among the first models applied in this domain.
The earliest generation-based coverless steganography approach was introduced in 2018 by Duan et al. \cite{duan2018coverless}, pioneering the direct creation of stego-images from secret images and marking a shift from modification-based techniques.
Subsequent improvements, such as the Deep Convolutional GANs (DCGANs) proposed by Hu et al. \cite{hu2018novel}, enhanced image quality and robustness against steganalysis.
However, GAN-based methods often struggle to balance security, embedding capacity, and image quality.

Motivated by the limitations of GANs, recent studies have explored alternative generative frameworks for steganography.
For example, StarGAN has been employed to enhance flexibility \cite{chen2020novel}, while gradient descent optimization has been incorporated to improve robustness \cite{peng2022robust}.
Additionally, the Glow model has been utilized to leverage bijective mappings between latent and image spaces \cite{zhou2022secret}.
Despite these advancements, GAN-based methods remain constrained by pseudo-key dependency, limited content diversity, and vulnerability to compression and steganalysis attacks.
These challenges underscore the need for alternative generative approaches, such as diffusion models, to address persistent security and robustness concerns.

To overcome these limitations, this study investigates the potential of diffusion models for coverless steganography.
Unlike GANs, which suffer from pseudo-key dependency and mode collapse, diffusion models employ a fundamentally different approach. Denoising Diffusion Probabilistic Models (DDPMs)\cite{ho2020denoising} iteratively transform random noise into structured data, offering high image fidelity and improved robustness.
However, the slow sampling process of DDPMs limits their practical applicability.
To address this issue, Denoising Diffusion Implicit Models (DDIMs)\cite{song2020denoising} introduce a deterministic sampling mechanism that enhances efficiency while maintaining high image quality.
These attributes make DDIMs particularly suitable for coverless steganography, where precise and reliable extraction of secret information is critical.

Recent research has explored diffusion models for embedding diverse types of information into images.
For instance, some methods integrate textual data with visual content\cite{10637346}, demonstrating the potential for multimodal steganography, while others, such as CRoSS\cite{yu2024cross}, use diffusion models to transform secret images into generated images. However, these approaches rely heavily on pseudo-keys (conditions involved in generation prcess) for extracting secret information from stego-object, which are tightly coupled to the embedded content and require frequent key negotiations for each communication session. These limitations reduce their practicality for real-world covert communication.

Building on these insights, we propose a diffusion-based coverless steganography approach that addresses key security and efficiency challenges while enhancing practical applicability.
Our approach achieves visual fidelity comparable to or superior to GAN-based methods\cite{dhariwal2021diffusionmodelsbeatgans}, which often struggle to preserve image quality.
To eliminate pseudo-key dependency, we integrate reversible data hiding (RDH) with chaotic encryption, introducing a real-key mechanism with an expanded key space.
This significantly enhances security and confidentiality with a single key negotiation for multiple sessions.
By combining these innovations, our method advances the practicality of diffusion-based steganography, offering a new perspective on secure and efficient covert communication.

Our contributions can be summarized as follows.
\begin{enumerate}
	\setlength{\itemsep}{3pt}
	\renewcommand{\labelenumi}{(\arabic{enumi})}
	
	\item \textbf{Diffusion-Based Framework for Coverless Steganography}: We propose a DDIM-driven coverless steganography approach that generates high-quality stego-images while addressing the limitations of GAN-based methods, such as pseudo-key dependency and content diversity constraints.
	
	\item \textbf{Enhanced Security and Efficient Key Negotiation}: By introducing a real-key mechanism with an expanded key space, our method eliminates reliance on pseudo-keys. It also reduces operational complexity by requiring only a single key negotiation for multiple communication sessions, improving practicality in secure communication scenarios.
	
	\item \textbf{Chaotic Encryption for Better Security}: To further enhance security, we incorporate chaotic encryption and RDH, minimizing correlations between the keys and secret images. This strengthens resistance against potential attacks.
	
	\item \textbf{Demonstrated Security and Practicality}: Experimental results validate that our approach resists substitution attacks while allowing receivers to authenticate the integrity of stego-objects.
	
\end{enumerate}

\section{Related Works}

\subsection{Generation-based coverless steganography}
Typical image steganography embeds secret information into pre-existing images by modifying their statistical or perceptual features.
Methods like Least Significant Bit (LSB) substitution are widely used but often introduce detectable statistical anomalies, compromising security.
Advanced adaptive steganography, such as HUGO\cite{HUGO}, WOW\cite{WOW}, and S-UNIWARD\cite{S-UNIWARD}, address this issue by selecting embedding positions based on the image content, thereby enhancing security.

In contrast, generation-based coverless steganography eliminates the need for modifying existing images by directly generating stego-images from secret information. Leveraging generative models like GANs and diffusion models, this approach maps secret data to image features, producing high-quality images that are statistically indistinguishable from natural images. This method not only enhances security and imperceptibility but also offers flexibility in generating diverse images tailored to various application scenarios.

\begin{figure}[h]
	\centering
	\includegraphics[width=0.75\linewidth]{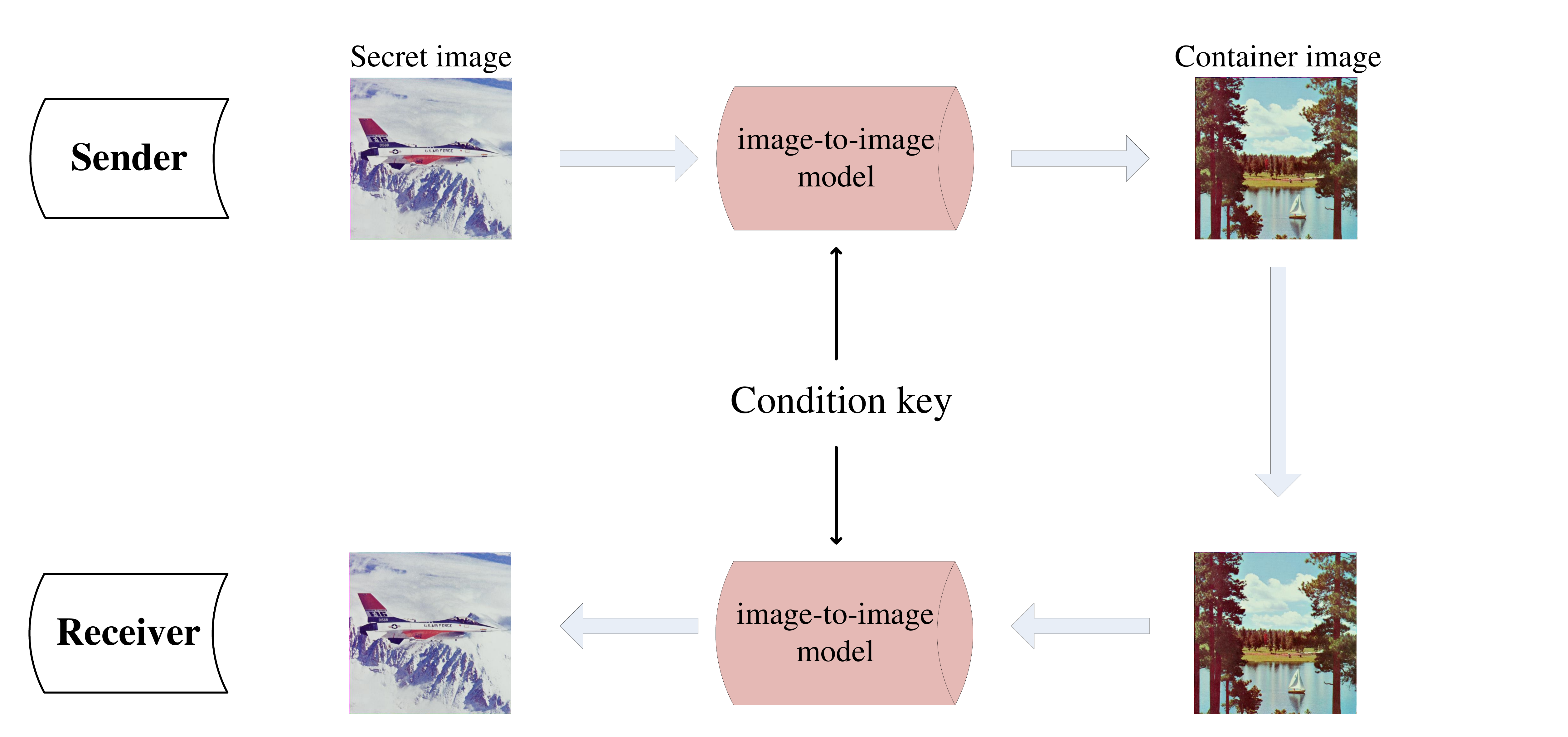}
	\caption{Framework of common generation-based coverless image steganography}
	\label{fig:framework_common_scheme}
\end{figure}

Specifically, in generation-based coverless steganography, the sender utilizes an image-to-image generative model, taking a secret image and a predefined key as inputs to generate a container image.
The receiver, equipped with the same key as agreed upon during the communication setup, extracts the secret image from the container image using the generative model’s reverse process.
This approach ensures security and concealment, as it avoids any direct modifications to existing images while enabling accurate reconstruction of the original secret image.
Figure \ref{fig:framework_common_scheme} illustrates the general framework of generation-based coverless steganography.

\subsection{Reversible data hiding}
Reversible Data Hiding (RDH), also known as lossless data embedding, has been extensively studied for its ability to embed secret information into a cover image while allowing complete recovery of the original image.
Traditional RDH methods, such as Difference Expansion (DE)\cite{tian2002wavelet,tian2003reversible}, Prediction Error Expansion (PEE)\cite{fallahpour2008reversible,hu2008based}, Histogram Shifting (HS)\cite{ni2006reversible,ying2019reversible}, and combined method of Prediction Error Histogram Shifting (HS-PEE)\cite{luo2009reversible,jia2019reversible}, introduced various strategies to minimize image distortion during embedding.
These techniques enable both high embedding capacity and reversibility, making them well-suited for applications where image integrity is crucial.

\begin{figure}[bth]
	\centering
	\includegraphics[width=0.7\linewidth]{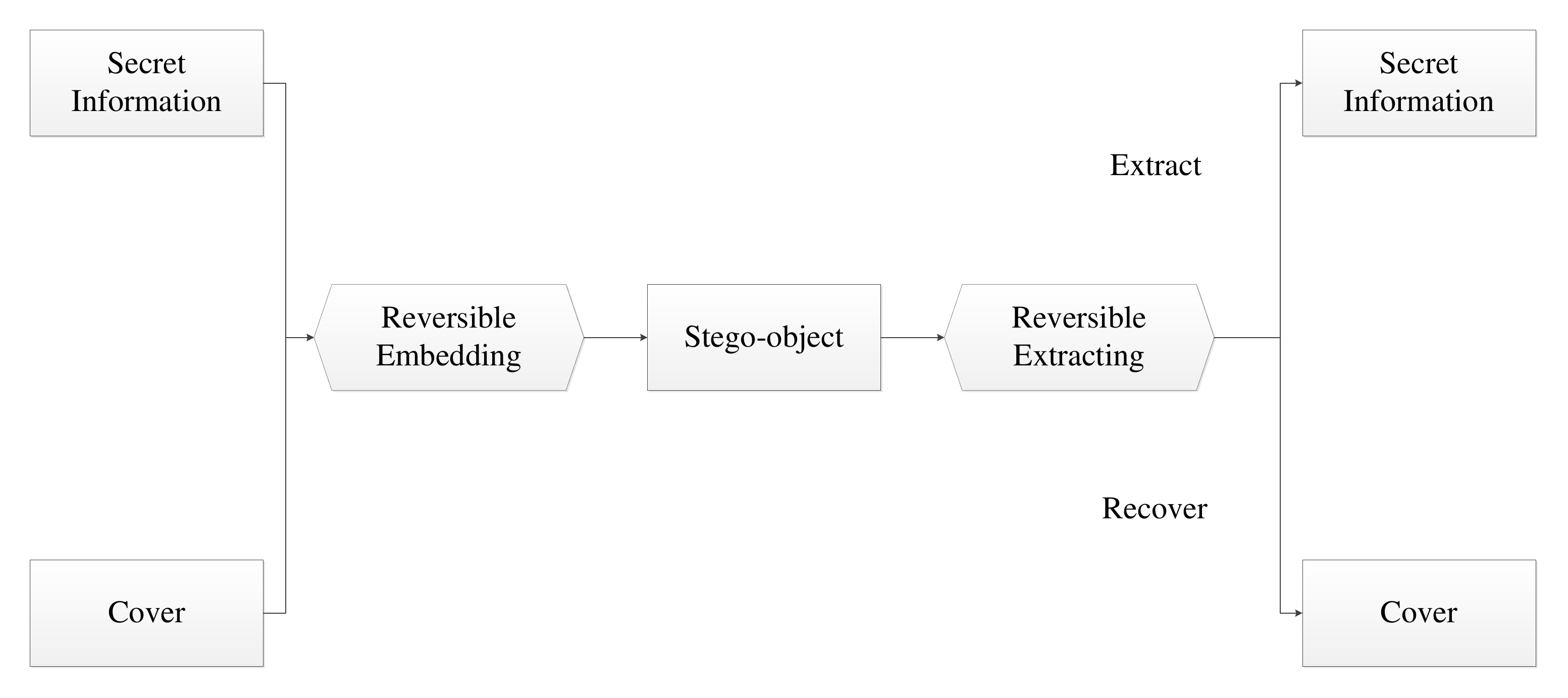}
	\caption{Communication model of reversible information hiding}
	\label{fig:model_rdh}
\end{figure}

The core idea of RDH is to exploit redundancy in the cover image, such as prediction errors or histogram characteristics, to create embedding space without introducing irreversible changes.
For instance, HS-based methods utilize the frequency of pixel values to shift the histogram and embed secret information into specific bins.
Similarly, DE-based approaches expand pixel differences to accommodate hidden data while preserving the original pixel relationships.

Figure \ref{fig:model_rdh} illustrates the general communication model for RDH.
The sender embeds secret information into the cover image using reversible embedding techniques, producing a stego-object.
At the receiver end, the extraction process retrieves the hidden information while restoring the original cover image without any loss.
These attributes make RDH a vital tool in secure communications and multimedia authentication.

\subsection{Diffusion Model Defined by DDIMs}
Diffusion models, such as Denoising Diffusion Probabilistic Models (DDPMs), represent a significant advancement in generative modeling by iteratively transforming random noise into structured data.
Despite their ability to produce high-quality samples, DDPMs often suffer from slow sampling due to the large number of required iterations.
To address this, Denoising Diffusion Implicit Models (DDIMs) were introduced, offering an efficient alternative with deterministic sampling.
This adjustment allows for faster sampling and precise control over the generation process, which enables the generation of identical outputs for the same initial conditions, enhancing reproducibility in applications.
The reverse process in DDIMs can be described as:
\begin{equation}
	x_{t-1}=\sqrt{\alpha _{t-1}}\frac{x_t-\sqrt{1-\alpha _t}\epsilon _{\theta}\left( x_t,t \right)}{\sqrt{\alpha _t}}+\sqrt{1-\alpha _{t-1}-\sigma _{t}^{2}}\cdot \epsilon _{\theta}\left( x_t,t \right) +\sigma _t\epsilon _t
	\label{eq:revp}
\end{equation}
where \(x_t\) represents the data (image) in step \( t\), \(\alpha_t\) is a predefined parameter controlling the level of noise added at each step or the information retained during the reverse process.
\(\varepsilon_{\theta}(x_t,t)\) is the  learnable noise estimator used to predict the noise component at time step \( t\).
\(\sigma_t\) is the noise control parameter.

When \(\sigma_t=0\), the reverse process becomes a deterministic reconstruction process, that is, a noise sample determines a generated image.
By transforming Eq.~\ref{eq:revp}, the inversion process can be expressed as:
\begin{equation}
	\frac{x_t}{\sqrt{\alpha _t}}=\frac{x_{t-1}}{\sqrt{\alpha _{t-1}}}+\left( \sqrt{\frac{1-\alpha _t}{\alpha _t}}-\sqrt{\frac{1-\alpha _{t-1}}{\alpha _{t-1}}} \right) \epsilon _{\theta}\left( x_t,t \right) 
	\label{eq:invp}
\end{equation}
Let \(\varDelta t=t-\left( t-1 \right) =1\), make a transform of Eq.~\ref{eq:invp}:
\begin{equation}
	\frac{x_{t-\varDelta t}}{\sqrt{\alpha _{t-\varDelta t}}}=\frac{x_t}{\sqrt{\alpha _t}}+\left( \sqrt{\frac{1-\alpha _{t-\varDelta t}}{\alpha _{t-\varDelta t}}}-\sqrt{\frac{1-\alpha _t}{\alpha _t}} \right) \epsilon _{\theta}\left( x_t,t \right) 
	\label{eq:invp2}
\end{equation}
DDIM simplifies the reverse process by reformulating it as an ordinary differential equation (ODE).
Using numerical integration methods such as Euler's method.
The resulting approximation equation of Eq.~\ref{eq:invp2} is:
\begin{equation}
	y\left( t-\varDelta t \right) =y\left( t \right) +\varDelta t\cdot f\left( t,y\left( t \right) \right) 
	\label{eq:eu}
\end{equation}
where \(y\left( t \right) =\frac{x_t}{\sqrt{\alpha _t}}\) and \(f\left( t,y\left( t \right) \right) =\epsilon _{\theta}\left( x_t,t \right)\).
Here \(\varDelta t\) is an approximation to make it close to the Euler integral.

Split the time step \([0,T]\) into intervals of length \(\varDelta t=t_{i+1}-t_i\).
At each time step, we can calculate Eq.~\ref{eq:eu} as follows:
\begin{equation}
	\frac{x_{t_{i+1}}}{\sqrt{\alpha _{t_{i+1}}}}=\frac{x_{t_i}}{\sqrt{\alpha _{t_i}}}+\left( \sqrt{\frac{1-\alpha _{t_{i+1}}}{\alpha _{t_{i+1}}}}-\sqrt{\frac{1-\alpha _{t_i}}{\alpha _{t_i}}} \right) \epsilon _{\theta}\left( x_{t_i},t_i \right) 
	\label{eq:eu2}
\end{equation}
where \(x_{t_{i+1}}\) is calculated based on the \(t_i\) and \(x_i\) of the current time step and the value of the model \((x_{t_i,t_i})\).
This allows for efficient sampling while maintaining high fidelity, as the sampling process is reduced to a subset of time steps instead of the full set \(T\).

In this study, DDIM's deterministic and efficient sampling process plays a crucial role in embedding and recovering secret information.
By leveraging its precise reconstruction capability, the proposed steganographic framework achieves high security and imperceptibility, making it suitable for real-world applications.

\section{Methodology}
\subsection{Analysis}
In generation-based coverless steganography, keys serve as a critical component in ensuring secure communication by regulating the embedding and extraction processes.
Typical cryptographic key systems, such as asymmetric cryptography, distinguish between a public key (\(K_{pub}\)) used for encryption and a private key (\(K_{pri}\)) used for decryption.
The security of such systems relies on the fundamental principle that \(K_{pub}\) can be freely distributed while \(K_{pri}\) is kept private and undisclosed.
Each communication endpoint independently holds its own key pair, and secure message exchange requires that the sender encrypts data using the recipient’s public key, which can then only be decrypted using the corresponding private key.

\begin{figure}[h]
	\centering
	\includegraphics[width=0.75\linewidth]{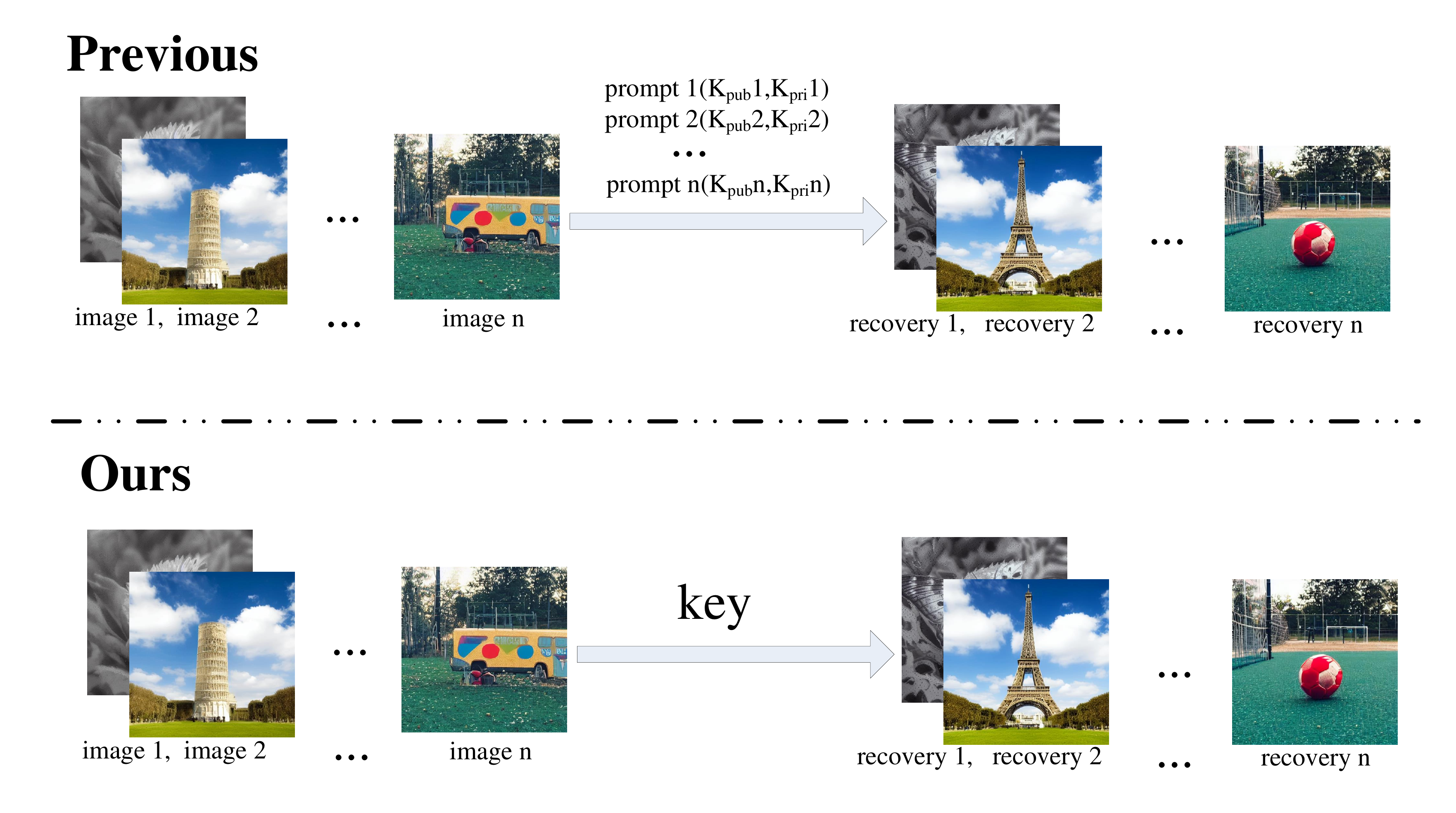}
	\caption{Comparison of previous and our method}
	\label{fig:comp_method_1}
\end{figure}

In contrast, existing generation-based coverless steganography schemes, such as CRoSS\cite{yu2024cross}, adopt a key system that differs fundamentally from asymmetric cryptography in their approach to key usage.
CRoSS utilizes a conditional diffusion model to generate stego-images, where both \(K_{pri}\) and \(K_{pub}\) function as symmetric keys and are required for both forward and reverse process.
Since these keys are directly linked to the generation process rather than acting as independent cryptographic entities, we refer to them as pseudo-keys.
This mechanism introduces several inherent limitations.
First, each secret image requires a distinct pseudo-key, leading to frequent key updates that cannot be controlled by the communicating parties.
Second, pseudo-keys impose a high transmission cost due to the large volume of exchanged key data.
Third, the strong correlation between pseudo-keys and secret images increases the risk of exposure if intercepted.
These limitations weaken key management efficiency, increase communication overhead, and undermine the security of covert communications.

The challenges posed by pseudo-keys in generation-based coverless steganography highlight the need for a more efficient key management scheme.
As discussed in \cite{bohme2010principles}, an optimal steganographic key system should balance security and efficiency to mitigate key exchange overhead and exposure risks.
To achieve this, we propose a real-key-based approach that replaces pseudo-keys with independently chosen cryptographic keys of fixed length, enhancing security while reducing transmission costs. Fig.~\ref{fig:comp_method_1} illustrates the comparison between pseudo-key-based methods and the proposed real-key approach.

\begin{figure}[bhtp]
	\centering
	\includegraphics[width=0.75\linewidth]{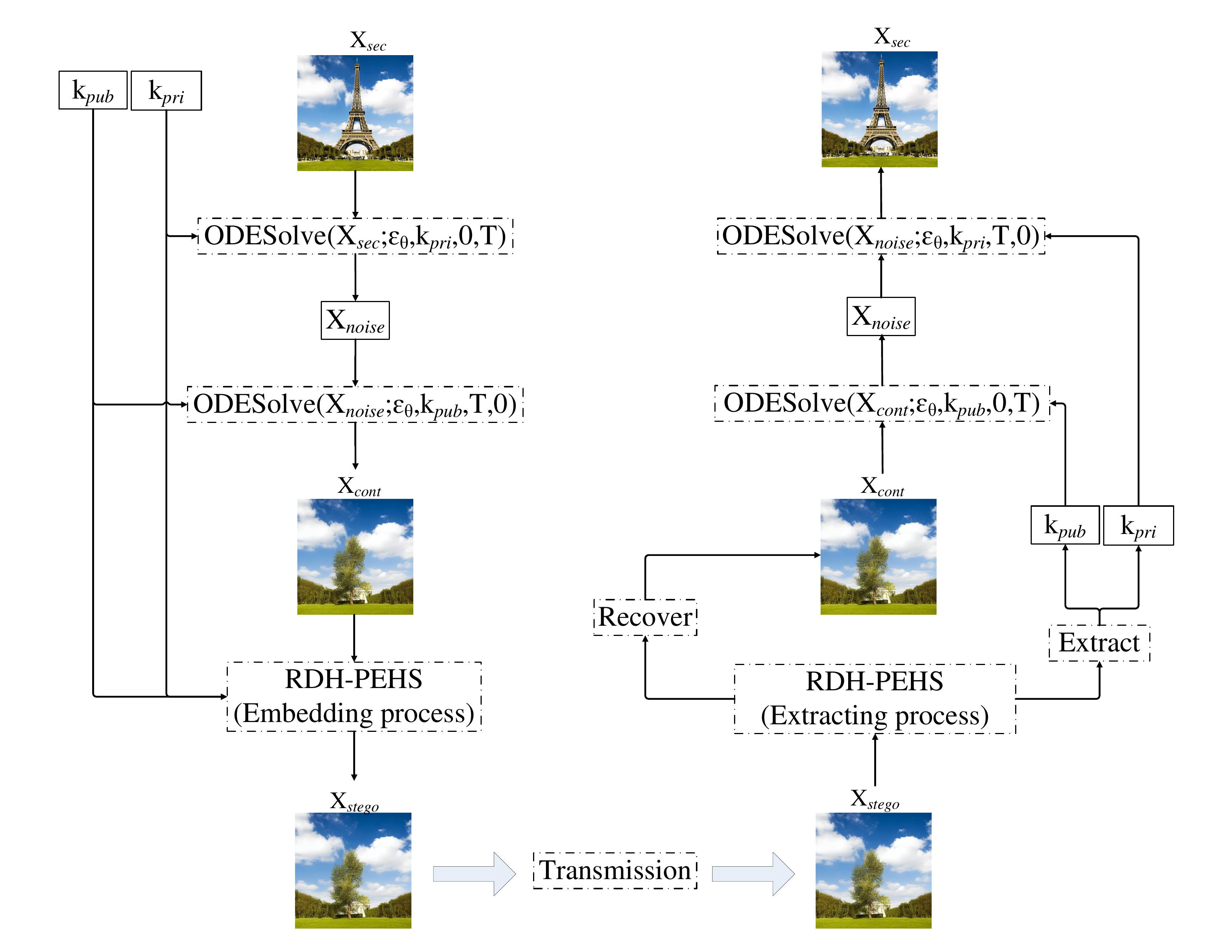}
	\caption{Framework of DDIM driven information hiding scheme without key}
	\label{fig:framework_without_key}
\end{figure}

\begin{figure}[bhtp]
	\centering
	\includegraphics[width=0.75\linewidth]{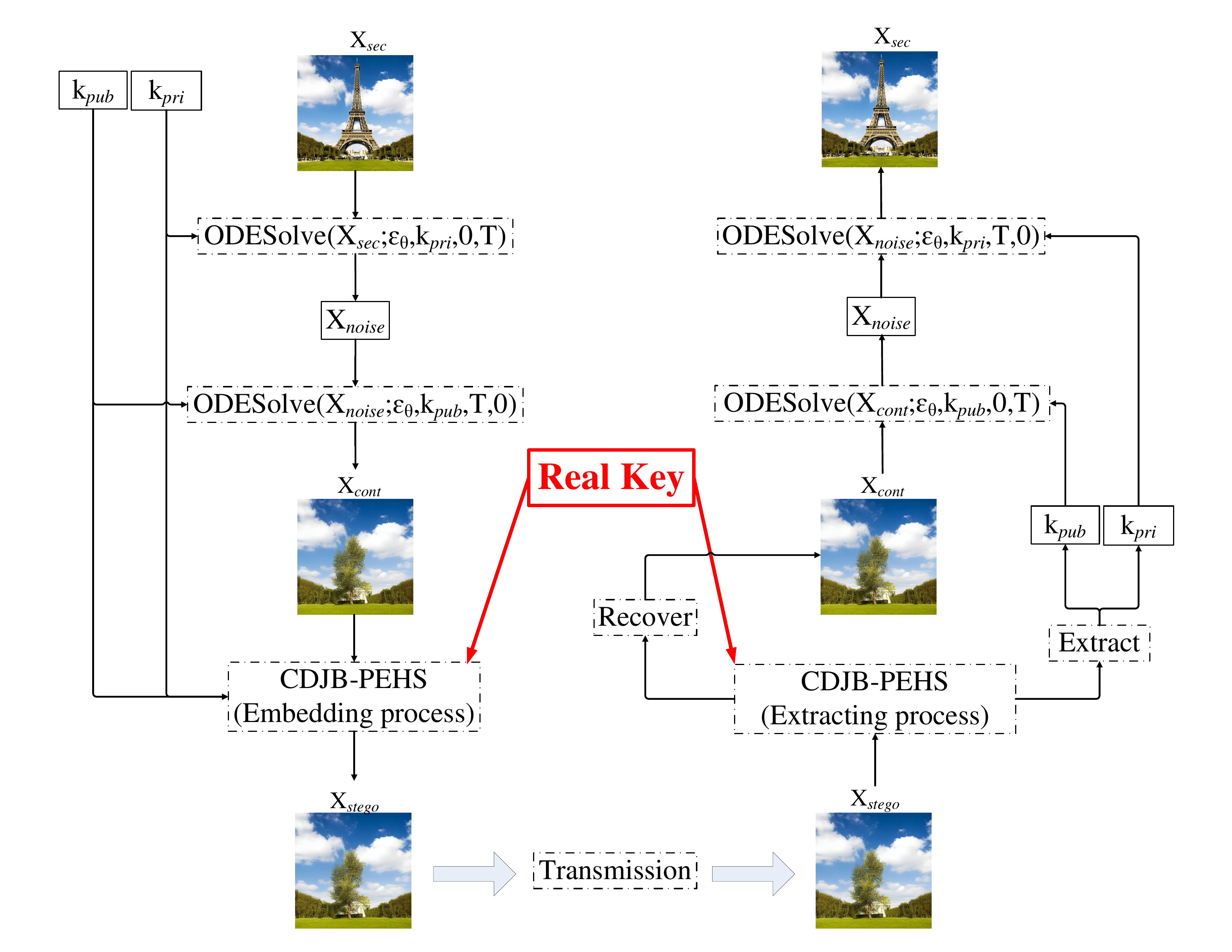}
	\caption{Framework of DDIM driven information hiding scheme with real key}
	\label{fig:framework_real_key}
\end{figure}

\subsection{Framework}
\label{sec:framework}

Designing a generation-based coverless steganography scheme requires balancing security, efficiency, and practical key management.
A critical consideration is how to avoid the constraints of pseudo-keys while ensuring reliable extraction and flexible information embedding.
In this work, we propose a DDIM-driven coverless steganography framework, which leverages Denoising Diffusion Implicit Models (DDIM) for structured image generation and integrates Reversible Data Hiding based on Prediction Error Histogram Shifting (RDH-PEHS) to embed auxiliary information without compromising imperceptibility.

The framework supports two configurations, each addressing different aspects of communication requirements.
The without-key scheme relies on DDIM’s deterministic sampling to reconstruct secret images without explicit key transmission, which simplifies deployment and lowers communication costs.
However, this approach may leave the system vulnerable to inference attacks when multiple stego-images are analyzed together.
To enhance security, the real-key scheme introduces fixed-length cryptographic keys that regulate the reconstruction process, preventing direct key-image associations and ensuring more stable key management.

\begin{algorithm}
	\caption{Without-key scheme}
	\label{alg:wo_key}
	\renewcommand{\algorithmicrequire}{\textbf{Input:}}
	\renewcommand{\algorithmicensure}{\textbf{Output:}}
	\begin{algorithmic}
		\Require Secret Image \(X_{sec}\), \(K_{pub}\), \(K_{pri}\)  \Comment{Sender}
		\State Initialize \textbf{DDIM} parameters
		\State \(X_{noise} \gets \text{ODESolve}(X_{sec};\varepsilon_{\theta},K_{pri},0,T)\) 
		\State \(X_{cont} \gets \text{ODESolve}(X_{noise};\varepsilon_{\theta},K_{pub},T,0)\) 
		\State \(X_{stego} \gets \text{RDH-PEHS}_{\text{Embedding}}(X_{cont}, K_{pub}, K_{pri}) \)
		\Ensure Stego Image \(X_{stego}\)
		\State Send \(X_{stego}\) to \textbf{Receiver} \Comment{Transmission}
		\Require Stego Image \(X_{stego}\) \Comment{Receiver}
		\State \(X_{cont}, K_{pub}, K_{pri} \gets \text{RDH-PEHS}_{\text{Extracting}}(X_{stego}) \)
		\State \(X_{noise} \gets \text{ODESolve}(X_{cont};\varepsilon_{\theta},K_{pub},0,T) \)
		\State \(X_{sec} \gets \text{ODESolve}(X_{noise};\varepsilon_{\theta},K_{pri},T,0) \)
		\Ensure Secret Image \(X_{sec}\)
	\end{algorithmic}
\end{algorithm}

\paragraph{Without-key scheme} As illustrated in Figure \ref{fig:framework_without_key}, the sender first applies the inversion process to transform the secret image \(X_{sec}\) into an intermediate noise representation \(X_{noise}\) with \(K_{pri}\) and then passed through the reverse process to generate the container image \(X_{cont}\) with \(K_{pub}\).
RDH-PEHS is then applied to \(X_{cont}\) to embed \(K_{pub}\) and \(K_{pri}\), producing the final stego-image \(X_{stego}\).

At the receiver side, the extraction process follows the inverse of the embedding phase.
RDH-PEHS is first used to extract \(K_{pub}\) and \(K_{pri}\) and recover the container image \(X_{cont}\) from setgo image \(X_{stego}\).
Subsequently, the inversion process reconstructs \(X_{noise}\) from \(X_{cont}\), followed by the reverse process to recover the secret image \(X_{sec}\).
Since the DDIM sampling is deterministic, a receiver with the same generative model can reconstruct the secret image.

However, since the mapping between noise and generated images is deterministic, this approach introduces potential security risks. An adversary with access to multiple stego-images may analyze these mappings to infer \(K_{pub}\) and \(K_{pri}\), making the scheme susceptible to inference attacks.

\begin{algorithm}
	\caption{Real-key scheme with CDJB-PEHS}
	\label{alg:wo_key}
	\renewcommand{\algorithmicrequire}{\textbf{Input:}}
	\renewcommand{\algorithmicensure}{\textbf{Output:}}
	\begin{algorithmic}
		\Require Secret Image \(X_{sec}\), \(K_{pub}\), \(K_{pri}\), \textbf{Real Key} \Comment{Sender}
		\State Initialize \textbf{DDIM} parameters
		\State \(X_{noise} \gets \text{ODESolve}(X_{sec};\varepsilon_{\theta},K_{pri},0,T)\) 
		\State \(X_{cont} \gets \text{ODESolve}(X_{noise};\varepsilon_{\theta},K_{pub},T,0)\) 
		\State \(X_{stego} \gets \text{CDJB-PEHS}_{\text{Embedding}}(X_{cont}, K_{pub}, K_{pri}, \textbf{Real Key}) \)
		\Ensure Stego Image \(X_{stego}\)
		\State Send \(X_{stego}\) to \textbf{Receiver} \Comment{Transmission}
		\Require Stego Image \(X_{stego}\), \textbf{Real Key} \Comment{Receiver}
		\State \(X_{cont}, K_{pub}, K_{pri} \gets \text{CJBD-PEHS}_{\text{Extracting}}(X_{stego}, \textbf{Real Key}) \)
		\State \(X_{noise} \gets \text{ODESolve}(X_{cont};\varepsilon_{\theta},K_{pub},0,T) \)
		\State \(X_{sec} \gets \text{ODESolve}(X_{noise};\varepsilon_{\theta},K_{pri},T,0) \)
		\Ensure Secret Image \(X_{sec}\)
	\end{algorithmic}
\end{algorithm}

\paragraph{Real-key scheme with CDJB-PEHS} To mitigate security risks associated with deterministic mappings, the key-based approach introduces a \textbf{Real Key}, a fixed-length cryptographic key that is independent of the generative process, to regulate the embedding positions of \(K_{pub}\) and \(K_{pri}\), as illustrated in Figure \ref{fig:framework_real_key}.
Compared to without-key scheme, this approach selectively determines embedding positions using Chaos-Driven Jump Bit-Prediction Error Histogram Shifting (CDJB-PEHS) instead of embedding information in sequence.
The real key, which can be securely transmitted through a protected channel or pre-shared between communicating parties, influences the chaotic sequence that dictates the embedding locations within the container image \( X_{cont} \), ensuring that the embedding pattern remains unpredictable.

At the receiver side, the extraction process mirrors the embedding procedure.
The same chaotic sequence, derived from the shared \textbf{Real Key}, is used to identify the embedding locations and extract the \(K_{pub}\) and \(K_{pri}\).
Once the stego-image is processed through CDJB-PEHS extraction, the inversion process reconstructs \( X_{noise} \) from \( X_{cont} \), followed by the reverse process to recover the secret image \( X_{sec} \).
Since the embedding locations are key-dependent, an adversary without knowledge of the correct real key cannot reconstruct the original mapping, enhancing resistance against inference attacks.

With this design, the proposed framework offers a practical and efficient approach for covert communications.

\subsection{RDH-PEHS}
\label{sec:rdh_pehs}

Reversible data hiding (RDH) is a key component of the proposed steganographic framework, enabling auxiliary information to be embedded while ensuring complete recovery of the original content.
To achieve this, we adopt RDH-PEHS (\textbf{R}eversible \textbf{D}ata \textbf{H}iding based \textbf{P}rediction \textbf{E}rror \textbf{H}istogram \textbf{S}hifting), which provides high embedding capacity without compromising imperceptibility.

\begin{figure}[h]
	\centering
	\includegraphics[width=0.7\linewidth]{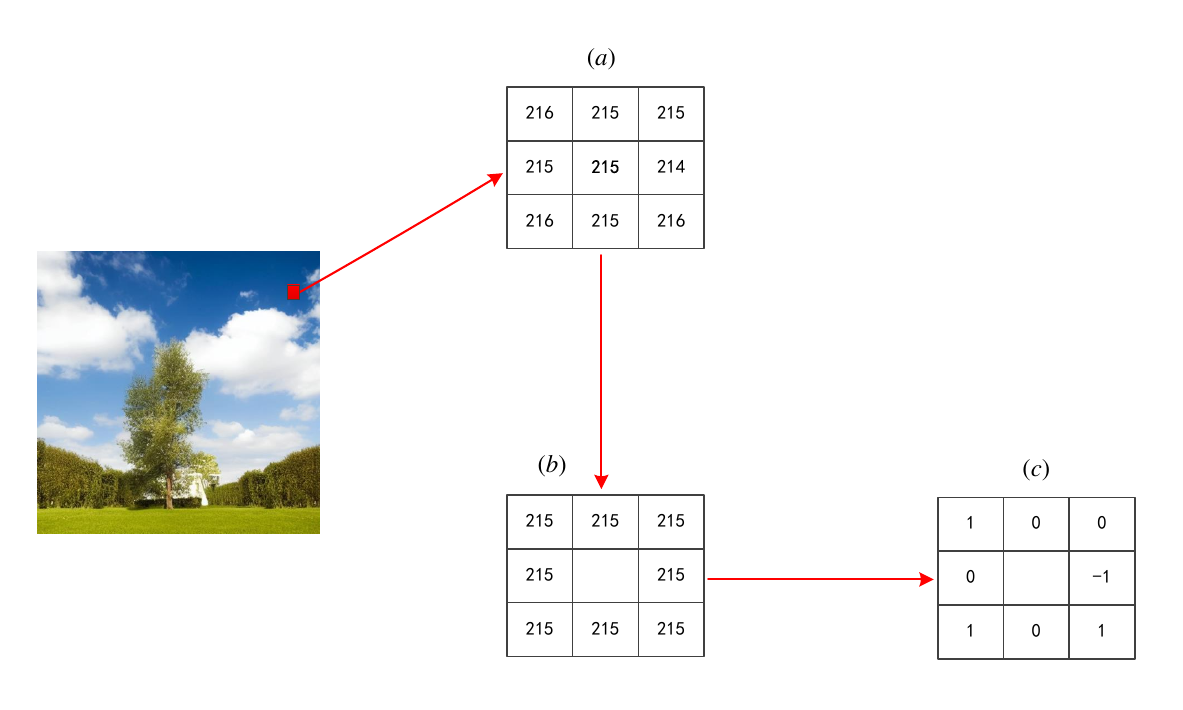}
	\caption{Process of Center Value Prediction}
	\label{fig:center_pre}
\end{figure}

\paragraph{Center Value Prediction}
The first step in RDH-PEHS is to compute the prediction error map, which forms the basis for reversible embedding.
Given an image of size \(M \times N\), it is divided into non-overlapping \(3 \times 3\) blocks, as illustrated in Fig.~\ref{fig:center_pre}(a).
If complete blocks cannot be formed, only fully divisible regions are considered, while the remaining pixels remain unchanged during the whole process.

Let \(x_{i,j}\) denote the grayscale value of a pixel, and \(x^{\prime}_{i,j}\) be the predicted value.
It is determined by the center pixel of the block, as shown in Fig.~\ref{fig:center_pre}(b):
\begin{equation}
	x^{\prime}_{i,j} = x_{2,2}, \quad (i,j) \neq (2,2)
\end{equation}
The prediction error is then computed as:
\begin{equation}
	e_{i,j} = x_{i,j} - x^{\prime}_{i,j}, \quad (i,j) \neq (2,2)
\end{equation}
where \(e_{i,j}\) represents the difference between the actual and predicted values. Fig.~\ref{fig:center_pre}(c) illustrates an example of the computed prediction error.

\paragraph{Embedding Process}
To embed secret bits, RDH-PEHS applies histogram shifting to create space in the prediction error histogram while preserving reversibility.
Experimental results indicate that the peak value of the histogram consistently occurs at zero\cite{xue2022thesis}, allowing \(p = 0\) to be directly assigned.
The first zero-frequency bin to the right of \(p\) is denoted as \(a\).

To illustrate the embedding process, we take right shifting as an example.
The histogram shifting process modifies values within the interval \((0, a)\) as follows:
\begin{equation}
	H(e+1) = H(e), \quad 0 < e < a
\end{equation}
\begin{align}
	e'_{i,j} =
	\begin{cases}
		e_{i,j} + 1, & 0 < e_{i,j} < a \\
		e_{i,j}, & \text{otherwise}
	\end{cases}
\end{align}
where \(H(e)\) represents the number of pixels with a prediction error value of \(e\), and \(H(a) = 0\) denotes the first zero-frequency bin.
For right shifting, the operation is performed sequentially from right to left, ensuring that each value is updated in a cascading manner, preventing data conflicts.

After shifting, secret bits are embedded into eligible pixels, defined as non-center pixels where \(e_{i,j} = 0\). The embedding rule is given by:
\begin{align}
	e'_{i,j} =
	\begin{cases}
		e_{i,j} + m, & e_{i,j} = 0, \quad m \in \{0,1\} \\
		e_{i,j}, & \text{otherwise}
	\end{cases}
\end{align}
where \(m\) represents the embedded bit.

The RDH-PEHS embedding process differs between the two configurations.
In the without-key scheme, all eligible pixels are used for embedding, ensuring maximal embedding capacity.
Conversely, in the real-key scheme, the embedding positions are selectively determined using a chaotic sequence, as described in Sec.~\ref{sec:cdjb_pehs}, restricting embedding to a subset of available locations, thereby increasing security.

\begin{figure}[h]
	\centering
	\includegraphics[width=0.8\linewidth]{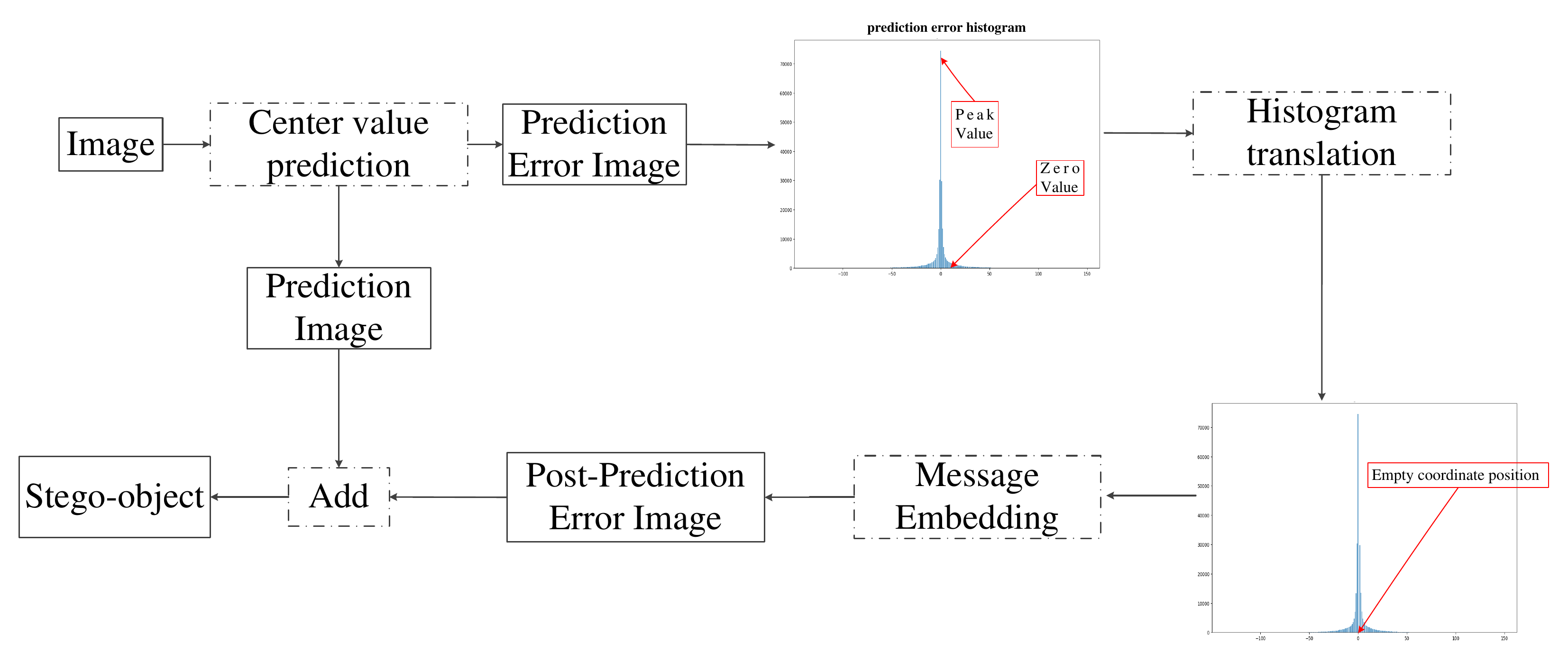}
	\caption{RDH-PEHS Embedding Process}
	\label{fig:rdh_emb_process}
\end{figure}

The detailed embedding workflow is illustrated in Fig.~\ref{fig:rdh_emb_process}.

\begin{figure}[ht]
	\centering
	\includegraphics[width=0.8\linewidth]{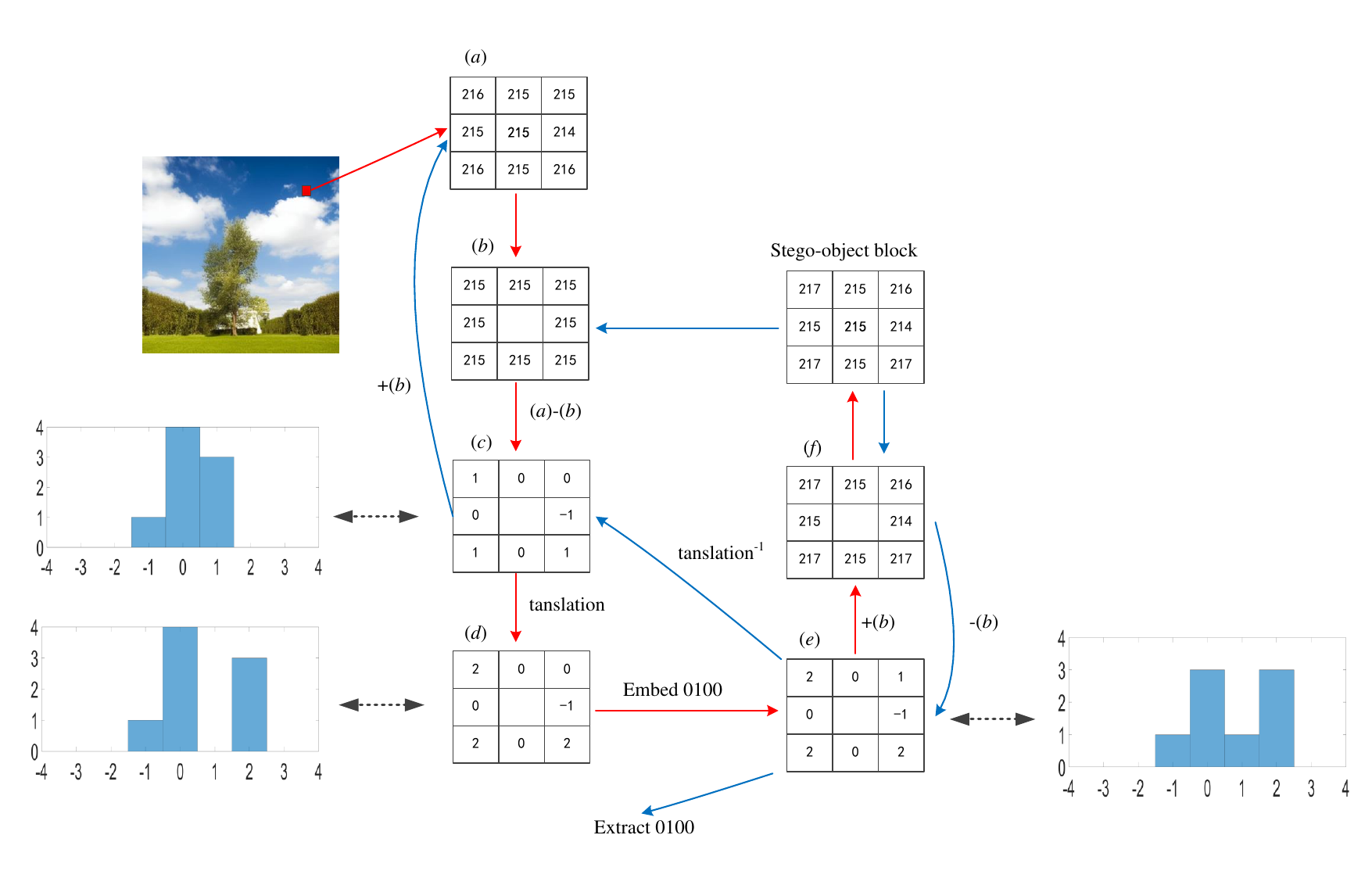}
	\caption{RDH-PEHS Process}
	\label{fig:rdh_process}
\end{figure}

\paragraph{Extracting Process}
To recover the embedded information and restore the original image, the receiver reverses the embedding process.
First, \textbf{Center Value Prediction} is applied to obtain the predicted values, generating an estimated prediction error map \(P_e\), similar to Fig.~\ref{fig:rdh_process}(b).
The stego-object block is processed by ignoring its center pixel, producing an intermediate block, as shown in Fig.~\ref{fig:rdh_process}(f).
The modified block is then subtracted from \(P_e\), yielding a prediction error map with embedded data, as illustrated in Fig.~\ref{fig:rdh_process}(e).

Next, the receiver extracts secret bits by scanning the prediction error map in a raster order (left to right, top to bottom).
The extraction rule is given by:
\begin{align}
	m =
	\begin{cases}
		1, & e'_{i,j} = 1, \quad \text{then set } e'_{i,j} \leftarrow 0 \\
		0, & e'_{i,j} = 0
	\end{cases}
\end{align}
The extracted bit sequence \(m\) represents the hidden information.

After extracting the embedded data, the receiver determines which values were shifted during embedding via the known zero-frequency bin \(a\).
The histogram shifting process is now performed in reverse order, moving values from \(H(e+1)\) back to \(H(e)\) for \(1 \leq e < a\):
\begin{equation}
	H(e) = H(e+1), \quad 1 \leq e < a
\end{equation}
\begin{align}
	e_{i,j} =
	\begin{cases}
		e'_{i,j} - 1, & 1 \leq e'_{i,j} < a \\
		e'_{i,j}, & \text{otherwise}
	\end{cases}
\end{align}
where the shifting operation is applied sequentially from left to right, ensuring that the original prediction error values are restored.

Finally, the recovered prediction error map is added back to the predicted values to reconstruct the original image:
\begin{equation}
	I_{\text{rec}} = P_e + e_{i,j}
\end{equation}
ensuring lossless recovery. The complete process is illustrated in Figure \ref{fig:rdh_process}.

\subsection{CDJB-PEHS}
\label{sec:cdjb_pehs}

Denoising Diffusion Implicit Models (DDIM) provide a deterministic generative process, making them suitable for coverless steganography.
The proposed scheme consists of two configurations: one that relies purely on DDIM’s deterministic mapping for secret image reconstruction and another that introduces a cryptographic key to enhance security. While the former avoids key management complexities, it is vulnerable to inference attacks, as adversaries with access to multiple communications may analyze deterministic mappings to reconstruct secret information.

CDJB-PEHS utilizes a piecewise logistic map\cite{ccelik2024hybrid} to generate a chaotic sequence, ensuring that the embedding positions vary unpredictably. Given the real key \((\mu, a_0)\), the chaotic sequence $a_n$ is generated as follows:
\begin{align}
	a_{n+1}=
	\begin{cases}
		4\mu a_n\left( 0.5-a_n \right), & 0<a_n<0.5\\
		1-4\mu a_n\left( 0.5-a_n \right) \left( 1-a_n \right), & 0.5\leq a_n<1
	\end{cases}
	\label{eq:7}
\end{align}
where $\mu \in (3.6,4)$ is a system parameter controlling the chaotic behavior, and \(a_0\in (0,1)\) initializes the pseudo-random sequence. Using this sequence, the embedding positions \(p_i\) in this work are dynamically determined as:
\begin{equation}
	p_i=p\times (i-1)+\lceil{5\times a_i}\rceil 
	\label{eq:8}
\end{equation}
This step ensures that each bit of embedded information is mapped to a dynamically determined position, preventing predictable patterns in the stego-object.

\begin{figure}[hbt]
	\centering
	\includegraphics[width=0.7\linewidth]{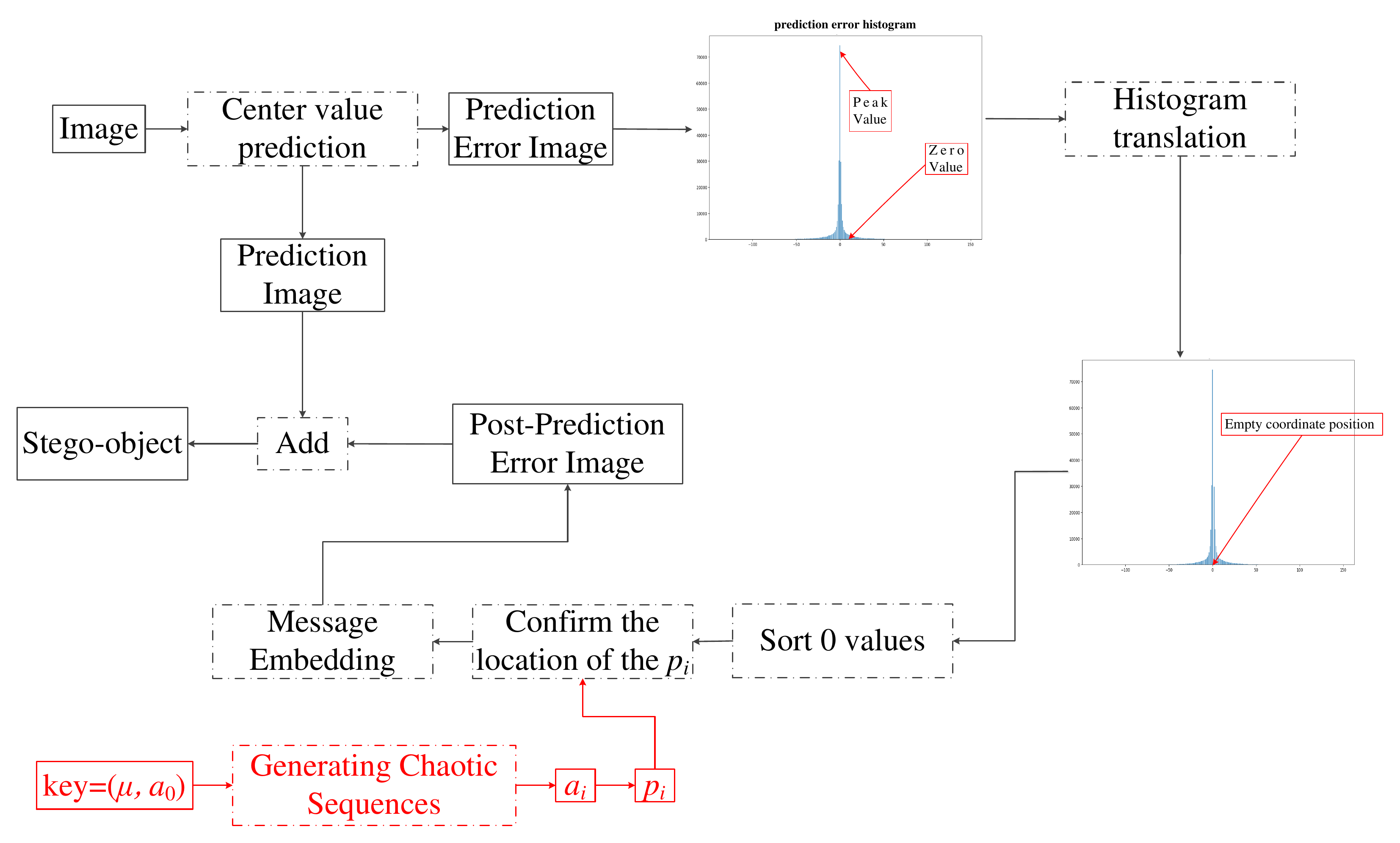}
	\caption{CDJB-PEHS Embedding Process}
	\label{fig:cdjb_emb_process}
\end{figure}

The embedding process follows the PEHS-based RDH strategy, where the prediction error histogram of the container image is modified to conceal secret information.
By incorporating the chaotic sequence into this process, the embedding locations are randomized, reducing the likelihood of structural inference attacks.
The embedding workflow of CDJB-PEHS is illustrated in Fig.~\ref{fig:cdjb_emb_process}.

At the receiver end, the same chaotic sequence, derived from the shared key, is used to identify the embedding positions and extract the secret information.
Since the key is independent of the container image, this approach eliminates the pseudo-key dependency observed in previous methods, where embedding conditions were directly linked to image content.
The extraction process mirrors the embedding phase, ensuring full reversibility.

By integrating a cryptographic key into the embedding process, the real-key scheme significantly enhances security while maintaining reversibility.
The use of a chaotic sequence ensures that embedding positions remain unpredictable, thereby mitigating structural analysis attacks. 
Table~\ref{tab:con_capa} demonstrates that DDIM-generated container images provide sufficient embedding capacity to support jump bit embedding.

\section{Experimental results and Analysis}

This section presents an experimental evaluation of the proposed steganographic scheme, focusing on key management efficiency, security, and embedding capacity.
The primary objective is to assess how the method reduces key transmission overhead, minimizes key-image correlation, and enhances authenticity against substitution attacks.

Existing diffusion-based coverless steganographic approaches typically rely on pseudo-keys derived from generation conditions, which necessitate frequent key exchanges and introduce strong correlations between the key and the secret information.
In contrast, the proposed method employs a real-key mechanism with a significantly reduced transmission complexity, ensuring secure communication while maintaining low correlation with the embeded information.
The integration of chaotic encryption further strengthens security by reducing predictability.

By addressing key transmission overhead, security risks associated with key-image dependency, and resilience to substitution attacks, the proposed scheme establishes a more secure and efficient framework for steganographic communication. 
The following sections provide a detailed quantitative assessment of these aspects.

\begin{table}[h]
	\centering
	\caption{Comparison of Key Management Strategies in Coverless Steganography}
	\renewcommand{\arraystretch}{1.5}
	\begin{tabular}{|c|c|c|}
		\hline
		\textbf{Aspect} & \textbf{Conventional} & \textbf{Proposed} \\ \hline
		Pre-negotiated Extraction Conditions & $n$ times & \textbf{1} time \\ \hline
		Modifications to Cover Image & No & No \\ \hline
		Additional Condition Transmission & Yes & \textbf{No} \\ \hline
	\end{tabular}
	\label{tab:key_management}
\end{table}

\begin{figure}[h]
	\centering
	\includegraphics[width=0.75\linewidth]{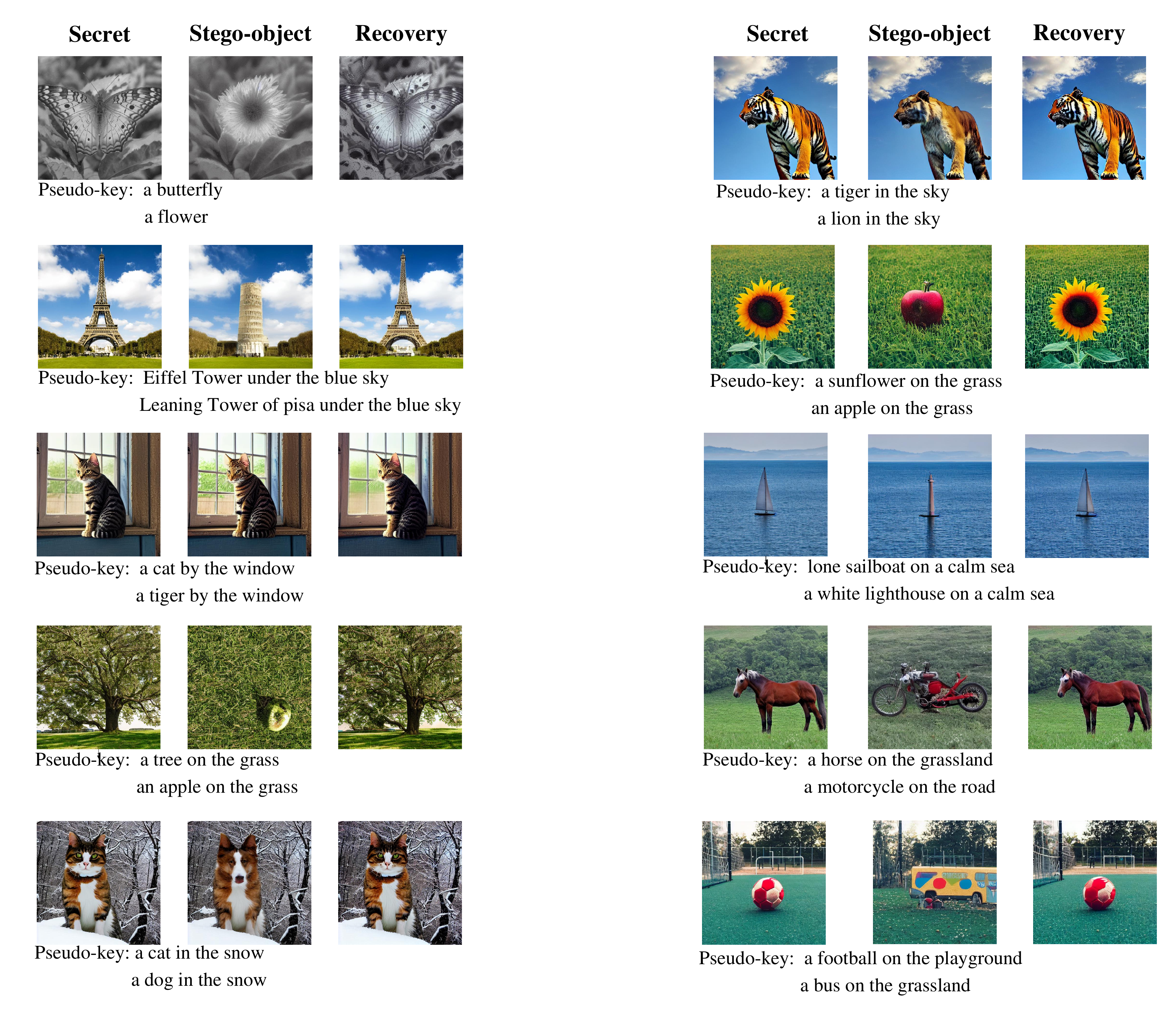}
	\caption{Pseudo-keys in typical coverless steganography}
	\label{fig:pkeys_in_typical}
\end{figure}

\subsection{Key Management}

Efficient key management is a fundamental requirement in steganographic communication.
Conventional generation-based coverless steganography relies on pseudo-keys that must be negotiated for each transmission session.
This process introduces a cumulative communication overhead of \(O(n)\) for \(n\) transmissions, increasing both key exchange complexity and the risk of exposure.

The proposed scheme addresses this limitation by establishing a single pre-negotiated key that remains valid across multiple sessions, reducing the transmission complexity to \(O(1)\).
Instead of transmitting a new extraction condition for each communication, our method embeds the necessary information directly within the generated container image using reversible data hiding.
As summarized in Table \ref{tab:key_management}, this eliminates the need for additional condition transmission while maintaining the integrity of the container image.

To quantify the reduction in key transmission overhead, we compare the number of bits required for secure key exchange.
For example, in the case of CRoSS \cite{yu2024cross}, the required bit of transmission of \textbf{10} secret images in Fig.~\ref{fig:pkeys_in_typical} is:

\begin{itemize}
	\item \textbf{Pseudo-key} Given that each secret image requires pseudo-key, resulting in a total overhead of \textbf{3568} bits.
	\item \textbf{Real-key (Ours)} A single pre-negotiated key is used across multiple transmissions, requiring only \textbf{102} bits in total, regardless of the number of images.
\end{itemize}

The value of \(\textbf{102}\) is determined by encoding the real-number parameters \(\mu\) and \(a_0\) in binary, where \(\mu\) is represented with 15 decimal places and \(a_0\) with 16 decimal places.
The first decimal digit of \(\mu\) has 4 possible values range in \([6,9]\) and the remaining 14 digits are unconstrained. 
According to Shannon’s information theory, the total required bit length is computed as \( \lceil \log_2(4 \times 10^{30}) \rceil =102\).

This drastic reduction minimizes the exposure of cryptographic parameters and significantly enhances the practicality of secure image communication.
The role of pseudo-keys in conventional methods is illustrated in Fig.~\ref{fig:pkeys_in_typical}, where each transmitted image requires a distinct key, contributing to the high transmission overhead.

This advantage is particularly relevant in scenarios requiring strict confidentiality.
For instance, in visual protection of classified information, such as secure government or military image transmissions, minimizing key exchange reduces the attack surface for adversaries attempting to intercept encryption parameters.
By ensuring that the key remains consistent across multiple communications, the proposed method mitigates potential security risks while maintaining operational efficiency.

\begin{figure}[ht]
	\centering
	\begin{subfigure}[b]{0.15\textwidth}
		\includegraphics[width=\textwidth]{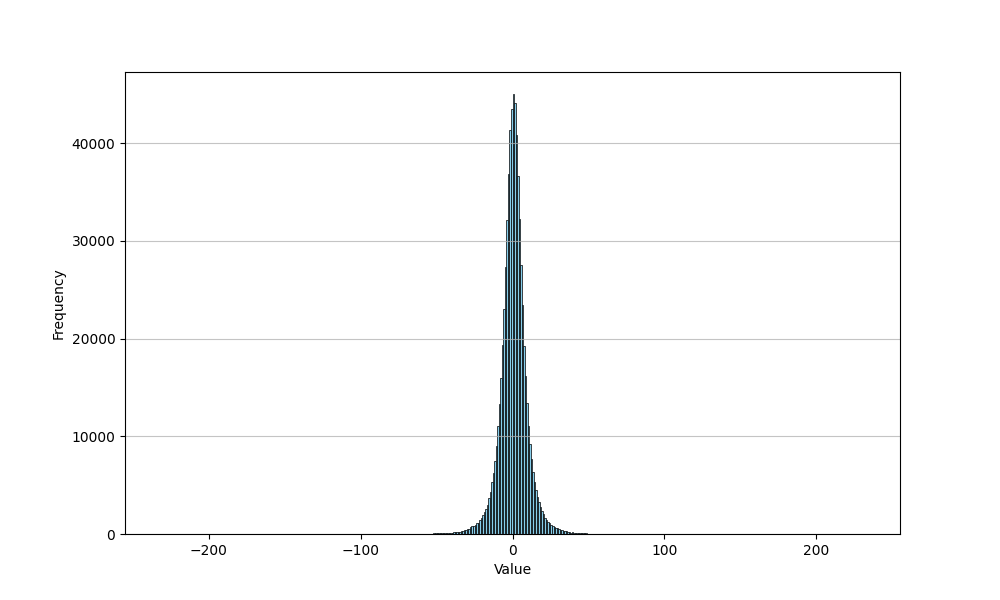}
		\caption{}
	\end{subfigure}
	\begin{subfigure}[b]{0.15\textwidth}
		\includegraphics[width=\textwidth]{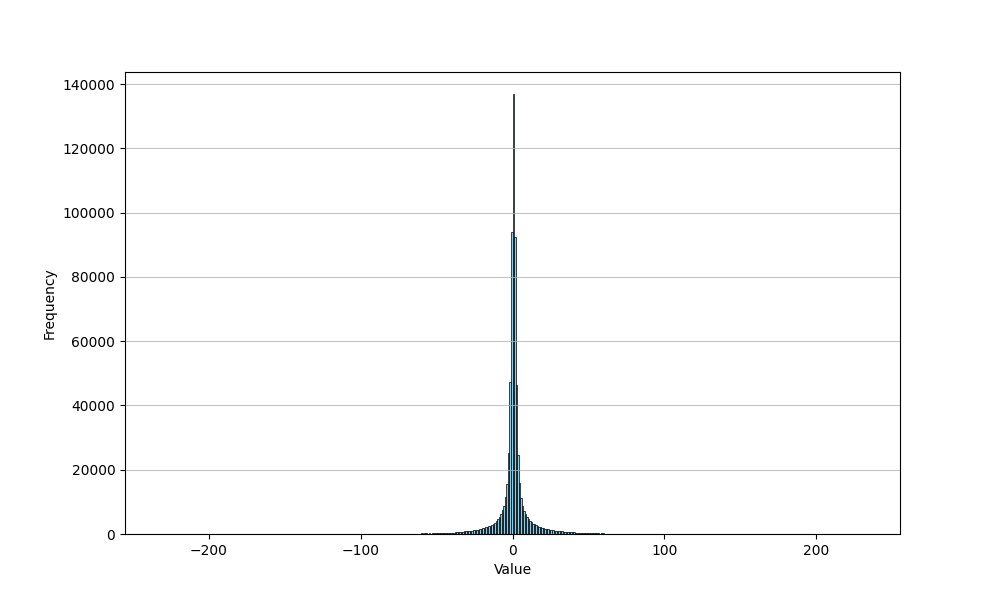}
		\caption{}
	\end{subfigure}
	\begin{subfigure}[b]{0.15\textwidth}
		\includegraphics[width=\textwidth]{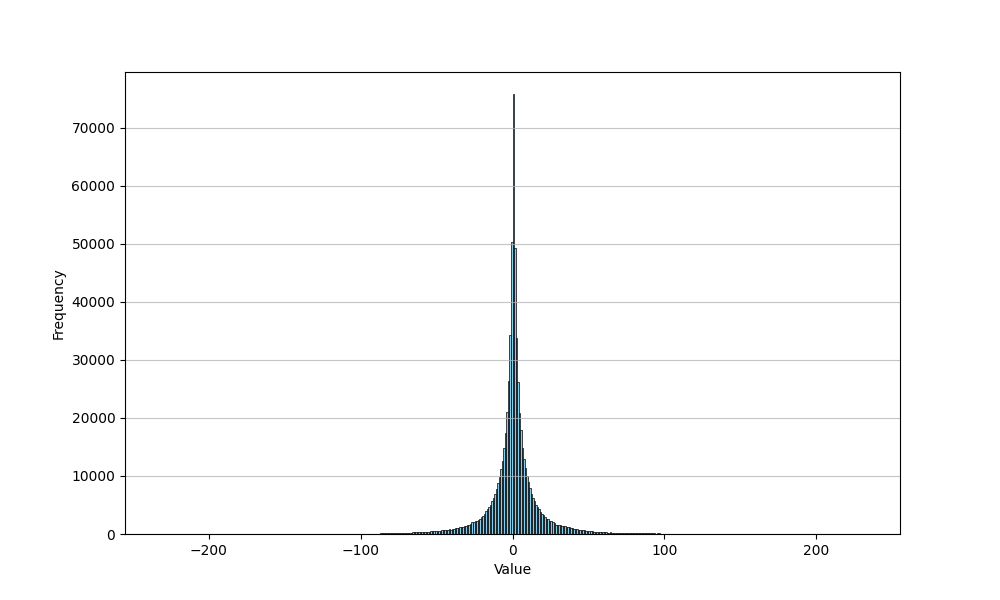}
		\caption{}
	\end{subfigure}
	\begin{subfigure}[b]{0.15\textwidth}
		\includegraphics[width=\textwidth]{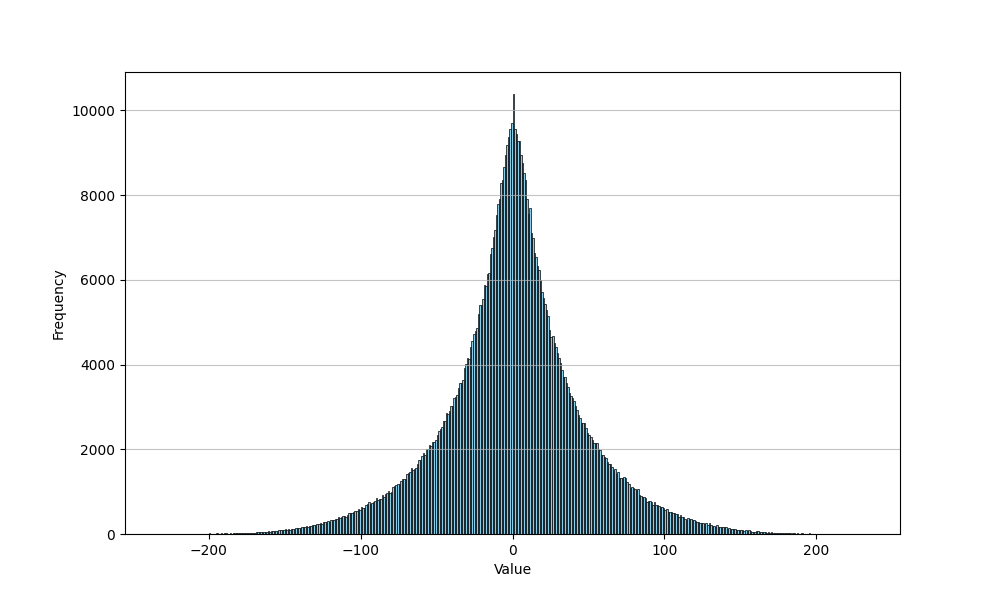}
		\caption{}
	\end{subfigure}
	\begin{subfigure}[b]{0.15\textwidth}
		\includegraphics[width=\textwidth]{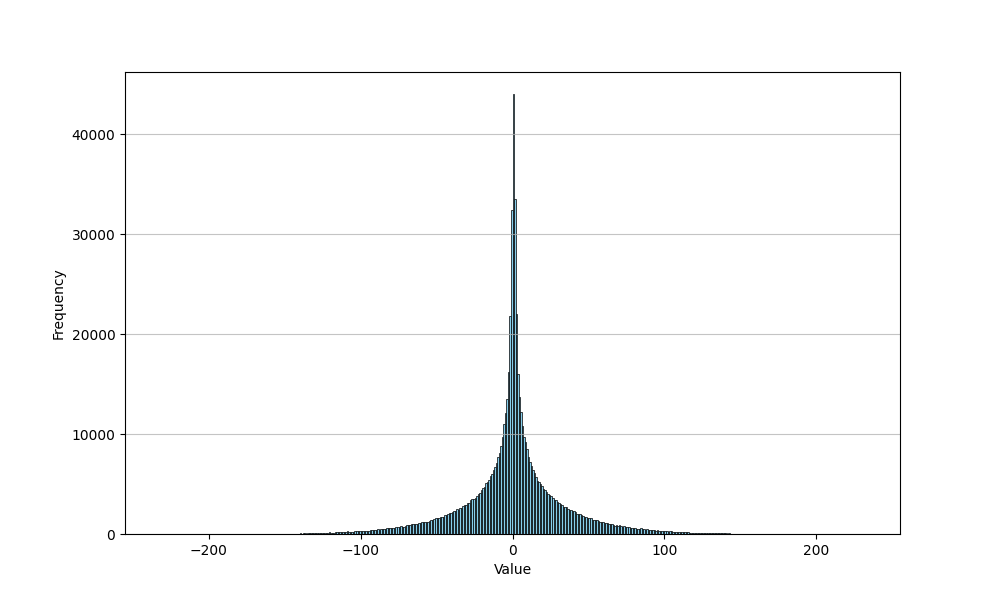}
		\caption{}
	\end{subfigure}
	
	\begin{subfigure}[b]{0.15\textwidth}
		\includegraphics[width=\textwidth]{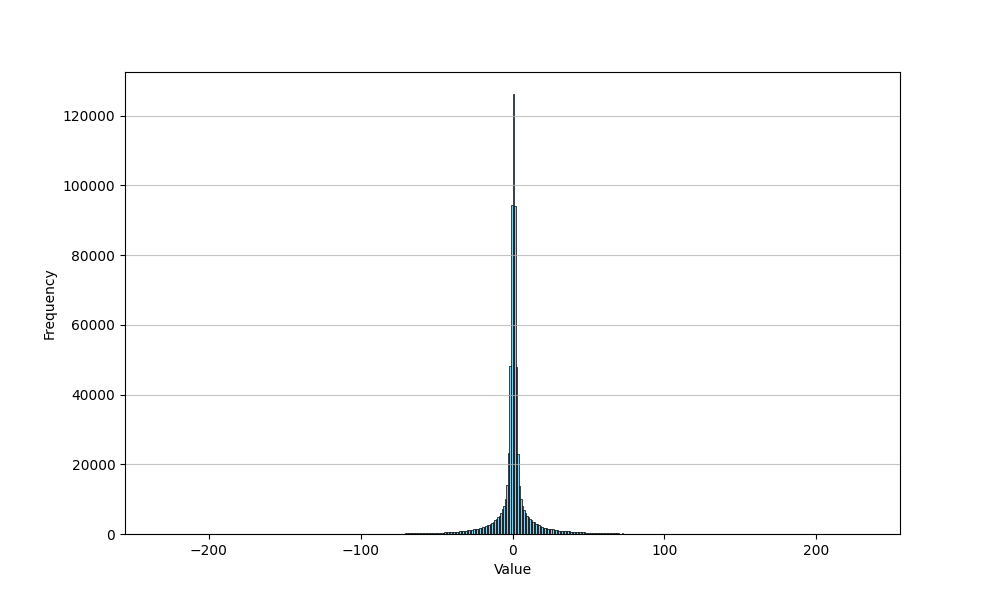}
		\caption{}
	\end{subfigure}
	\begin{subfigure}[b]{0.15\textwidth}
		\includegraphics[width=\textwidth]{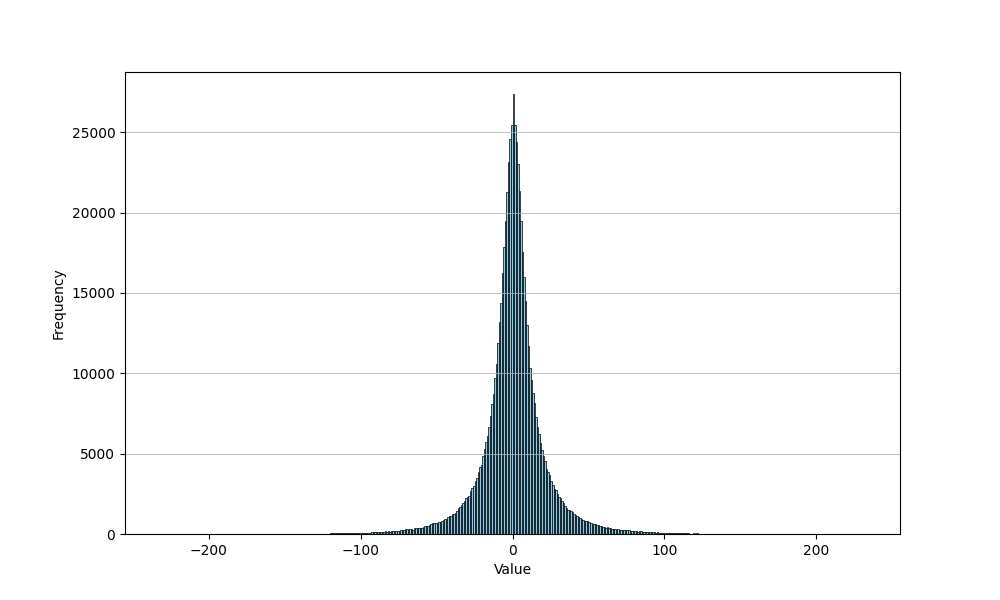}
		\caption{}
	\end{subfigure}
	\begin{subfigure}[b]{0.15\textwidth}
		\includegraphics[width=\textwidth]{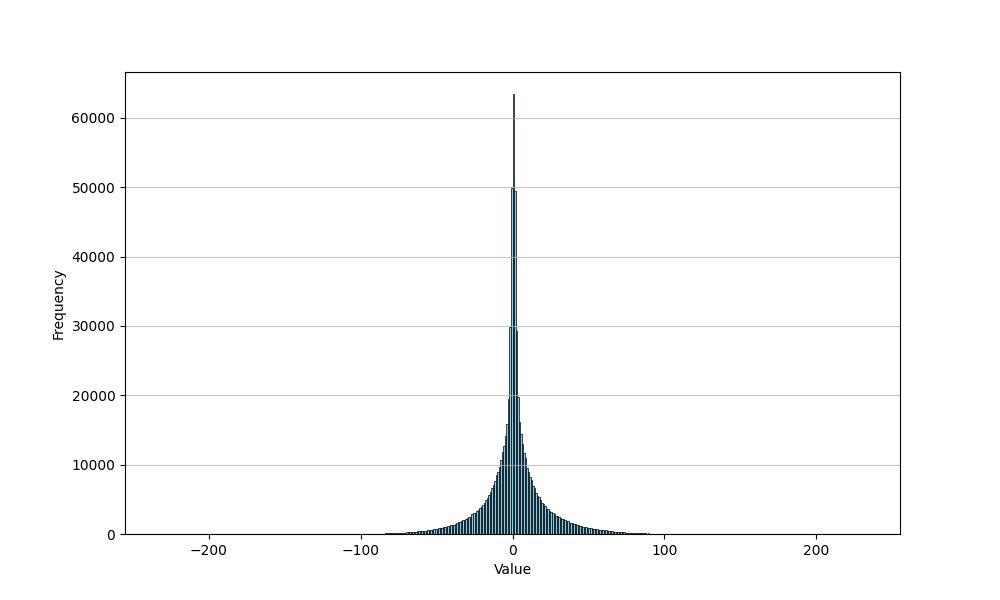}
		\caption{}
	\end{subfigure}
	\begin{subfigure}[b]{0.15\textwidth}
		\includegraphics[width=\textwidth]{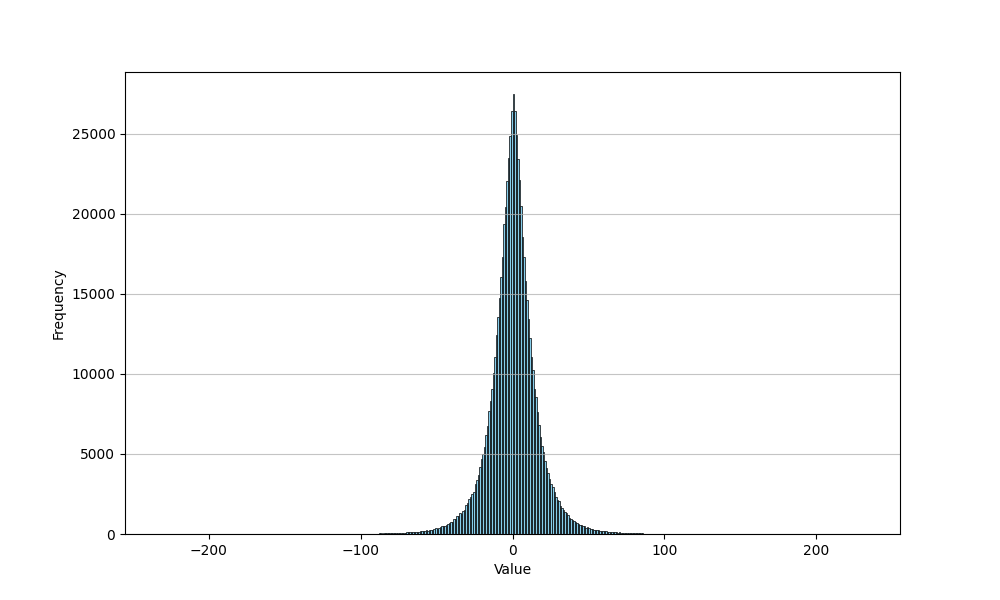}
		\caption{}
	\end{subfigure}
	\begin{subfigure}[b]{0.15\textwidth}
		\includegraphics[width=\textwidth]{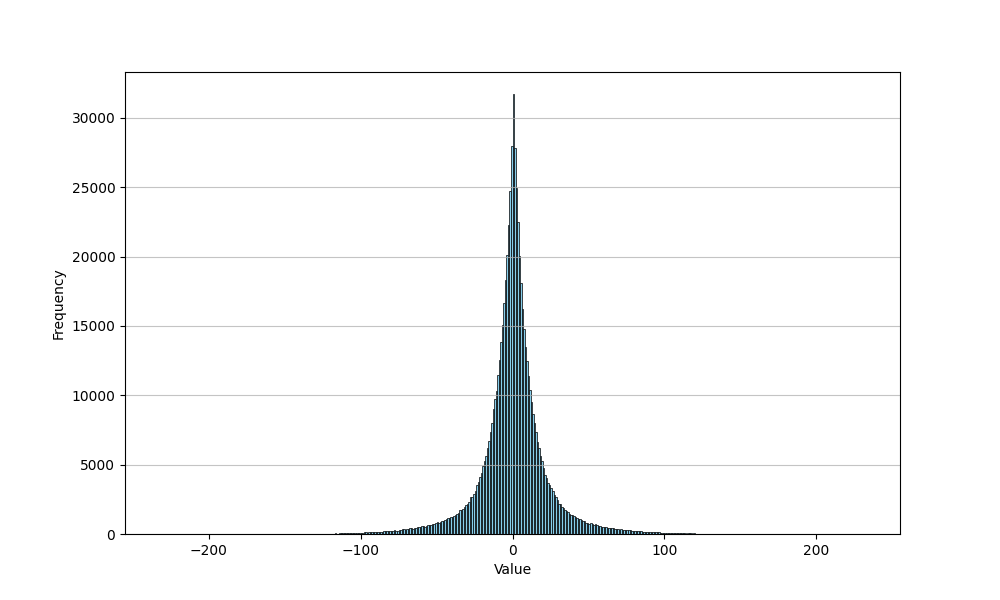}
		\caption{}
	\end{subfigure}
	\caption{Prediction error histogram for container images}
	\label{fig:peh_cont}
\end{figure}

\begin{table}[htb]
	\caption{Analysis of container image embedding capacity}
	\centering
	\begin{tabularx}{\linewidth}{
			>{\centering\arraybackslash}p{0.10\linewidth}
			>{\centering\arraybackslash}p{0.15\linewidth}
			>{\centering\arraybackslash}p{0.35\linewidth}
			>{\centering\arraybackslash}p{0.15\linewidth}
			>{\centering\arraybackslash}p{0.15\linewidth}
		}
		\toprule
		Image     & Capacity (bit) & Embedded Pseudo-key & Embedded (bit) & Required (bit)\\
		\midrule
		\multirow{2}*{(a)} & \multirow{2}*{45027} & a butterfly & \multirow{2}*{128} & \multirow{2}*{640} \\
		& & a flower \\
		\midrule
		\multirow{2}*{(b)} & \multirow{2}*{136913} & Eiffel Tower under the blue sky & \multirow{2}*{464} & \multirow{2}*{2320} \\
		& & Leaning Tower of Pisa under the blue sky \\
		\midrule
		\multirow{2}*{(c)} & \multirow{2}*{75777} & a cat by the window & \multirow{2}*{272} & \multirow{2}*{1360} \\
		& & a tiger by the window \\
		\midrule
		\multirow{2}*{(d)} & \multirow{2}*{10388} & a tree on the grass & \multirow{2}*{256} & \multirow{2}*{1280} \\
		& & an apple on the grass \\
		\midrule
		\multirow{2}*{(e)} & \multirow{2}*{44022} & a cat in the snow & \multirow{2}*{240} & \multirow{2}*{1200} \\
		& & a dog in the snow \\
		\midrule
		\multirow{2}*{(f)} & \multirow{2}*{126231} & a tiger in the sky & \multirow{2}*{256} & \multirow{2}*{1280} \\
		& & a lion in the sky \\
		\midrule
		\multirow{2}*{(g)} & \multirow{2}*{27379} & a sunflower on the grass & \multirow{2}*{288} & \multirow{2}*{1440} \\
		& & an apple on the grass \\
		\midrule
		\multirow{2}*{(h)} & \multirow{2}*{63461} & lone sailboat on a calm sea & \multirow{2}*{544} & \multirow{2}*{2720} \\
		& & a white lighthouse on a calm sea \\
		\midrule
		\multirow{2}*{(i)} & \multirow{2}*{27487} & a horse on the grassland & \multirow{2}*{416} & \multirow{2}*{2080} \\
		& & a motorcycle on the road \\
		\midrule
		\multirow{2}*{(j)} & \multirow{2}*{31716} & a football on the playground & \multirow{2}*{448} & \multirow{2}*{2240} \\
		& & a bus on the grassland \\
		\bottomrule
	\end{tabularx}
	\label{tab:con_capa}
\end{table}

\subsection{Embedding Capacity and Security}
\label{sec:emb_and_sec}

A practical steganographic system must balance embedding capacity and security.
While sufficient embedding capacity ensures that hidden information can be  transmitted, security considerations such as key-image correlation determine the system’s resilience against adversarial inference.

Since the proposed method employs CDJB-PEHS to embed  information, it is essential to evaluate whether the generated container images can provide adequate embedding capacity.
The CDJB-PEHS scheme introduces a constraint: for every five embedded bits, only one bit is effectively used for information, with the rest allocated for key-dependent encoding.
This significantly increases the required embedding capacity compared to direct data hiding techniques.

To assess the feasibility of this approach, we analyze the number of bits that can be embedded in container images generated by DDIM models.
As illustrated in Fig.~\ref{fig:peh_cont}, the prediction error histogram reveals that the value distribution is suited for histogram shifting, allowing efficient bit allocation without excessive distortion.
In the analyzed container images, the embedded pseudo-key components are separated using the \(\#\) symbol, which occupies a fixed 8-bit space per occurrence.
The quantitative evaluation in Table~\ref{tab:con_capa} further confirms that container images possess ample capacity to accommodate the required embeded information.

\begin{figure}[hbt]
	\centering
	\begin{subfigure}[b]{0.15\textwidth}
		\includegraphics[width=\textwidth]{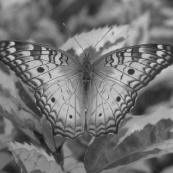}
		\caption{}
	\end{subfigure}
	\begin{subfigure}[b]{0.15\textwidth}
		\includegraphics[width=\textwidth]{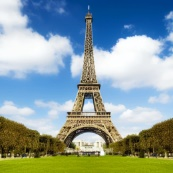}
		\caption{}
	\end{subfigure}
	\begin{subfigure}[b]{0.15\textwidth}
		\includegraphics[width=\textwidth]{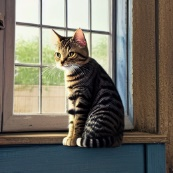}
		\caption{}
	\end{subfigure}
	\begin{subfigure}[b]{0.15\textwidth}
		\includegraphics[width=\textwidth]{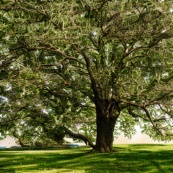}
		\caption{}
	\end{subfigure}
	\begin{subfigure}[b]{0.15\textwidth}
		\includegraphics[width=\textwidth]{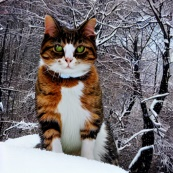}
		\caption{}
	\end{subfigure}
	
	\begin{subfigure}[b]{0.15\textwidth}
		\includegraphics[width=\textwidth]{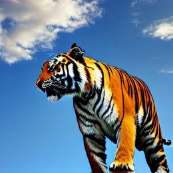}
		\caption{}
	\end{subfigure}
	\begin{subfigure}[b]{0.15\textwidth}
		\includegraphics[width=\textwidth]{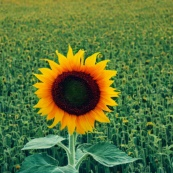}
		\caption{}
	\end{subfigure}
	\begin{subfigure}[b]{0.15\textwidth}
		\includegraphics[width=\textwidth]{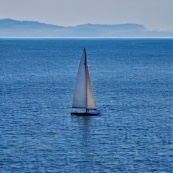}
		\caption{}
	\end{subfigure}
	\begin{subfigure}[b]{0.15\textwidth}
		\includegraphics[width=\textwidth]{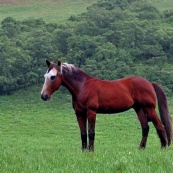}
		\caption{}
	\end{subfigure}
	\begin{subfigure}[b]{0.15\textwidth}
		\includegraphics[width=\textwidth]{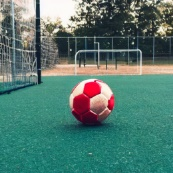}
		\caption{}
	\end{subfigure}
	\caption{Secret images}
	\label{fig:sec_img}
\end{figure}

\begin{table}[h]
	\caption{Comparison of Key and Secret Image Similarity (Sim.P = Similarity of Pseudo-key, Sim.R = Similarity of Real-key)}
	\centering
	\begin{tabular}{*{4}{c}}
		\toprule
		Image     & Pseudo-key & Sim.P & Sim.R\\
		\midrule
		\multirow{2}*{(a)} & a butterfly & 24.2892 & \multirow{2}*{17.0555} \\
		& a flower & 18.4283 &\\
		\midrule
		\multirow{2}*{(b)} & Eiffel Tower under the blue sky & 23.1538 & \multirow{2}*{14.4206} \\
		& Leaning Tower of Pisa under the blue sky & 15.0347 & \\
		\midrule
		\multirow{2}*{(c)} & a cat by the window & 27.1546 & \multirow{2}*{12.5790} \\
		& a tiger by the window & 24.6461 & \\
		\midrule
		\multirow{2}*{(d)} & a tree on the grass & 22.2781 & \multirow{2}*{13.2996} \\
		& an apple on the grass & 16.9600 & \\
		\midrule
		\multirow{2}*{(e)} & a cat in the snow & 27.1222 & \multirow{2}*{12.4728} \\
		& a dog in the snow & 18.5833 & \\
		\midrule
		\multirow{2}*{(f)} & a tiger in the sky & 25.0057 & \multirow{2}*{14.3805} \\
		& a lion in the sky & 19.3418 & \\
		\midrule
		\multirow{2}*{(g)} & a sunflower on the grass & 24.5156 & \multirow{2}*{13.0231} \\
		& an apple on the grass & 17.2146 & \\
		\midrule
		\multirow{2}*{(h)} & lone sailboat on a calm sea & 27.3972 & \multirow{2}*{14.2648} \\
		& a white lighthouse on a calm sea & 21.4917 & \\
		\midrule
		\multirow{2}*{(i)} & a horse on the grassland & 27.7675 & \multirow{2}*{13.5793} \\
		& a motorcycle on the road & 12.4741 & \\
		\midrule
		\multirow{2}*{(j)} & a football on the playground & 23.6410 & \multirow{2}*{12.9123} \\
		& a bus on the grassland & 12.9058 & \\
		\bottomrule
	\end{tabular}
	\label{tab:similarity}
\end{table}

Beyond embedding feasibility, the security of the steganographic system depends on the correlation between the key and the secret image.
A strong correlation increases the risk of information leakage, as an exposed key may inadvertently reveal details about the hidden content.
This risk is particularly evident in generation-based coverless steganography, where pseudo-keys often exhibit dependencies with the secret image, making them vulnerable to inference attacks.

By employing a real-key mechanism independent of the secret image, our method effectively reduces key-image correlation and mitigates the risk of information inference through key exposure.
Specifically, under the chaotic encryption configuration of \((\mu = 3.799200023214331,a_0 = 0.8888564633215454)\), the proposed approach performs steganographic embedding on the secret images in Fig.~\ref{fig:sec_img}, producing the similarity comparison results summarized in Table~\ref{tab:similarity}.
The CLIP-based \cite{hafner2021clip} evaluation demonstrates that pseudo-keys exhibit a high similarity to the corresponding secret images, whereas real-keys remain largely uncorrelated, ensuring stronger protection against key-based attacks.

These properties are particularly beneficial in privacy-preserving communications.
In identity anonymization, where sensitive images such as facial photographs must be transformed into non-identifiable cover images, reducing key-image correlation ensures that key exposure does not compromise security.
At the same time, the high embedding capacity allows for secure transmission of authentication metadata or additional security markers without perceptible image degradation.
This balance between capacity and security ensures that covert communication remains both imperceptible and resilient against potential attacks.

\begin{figure}[hb]
	\centering
	\begin{subfigure}[b]{0.15\textwidth}
		\includegraphics[width=\textwidth]{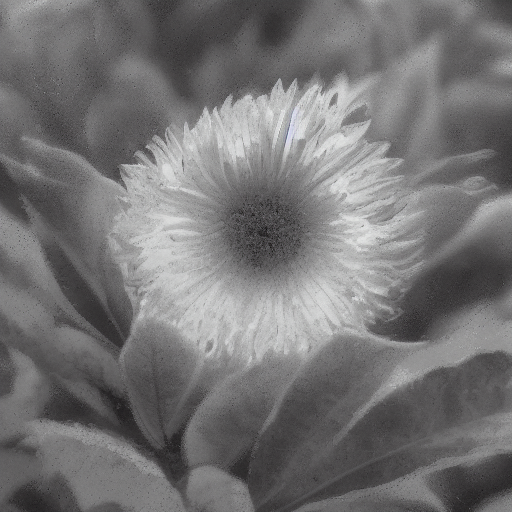}
		\caption{}
	\end{subfigure}
	\begin{subfigure}[b]{0.15\textwidth}
		\includegraphics[width=\textwidth]{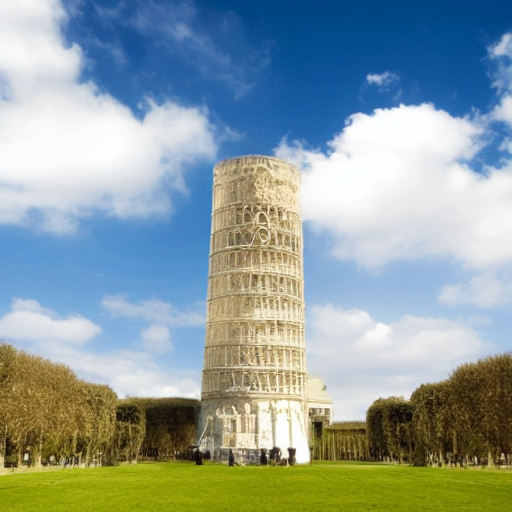}
		\caption{}
	\end{subfigure}
	\begin{subfigure}[b]{0.15\textwidth}
		\includegraphics[width=\textwidth]{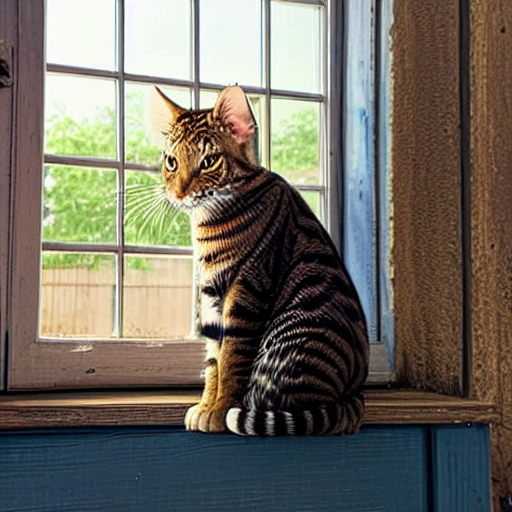}
		\caption{}
	\end{subfigure}
	\begin{subfigure}[b]{0.15\textwidth}
		\includegraphics[width=\textwidth]{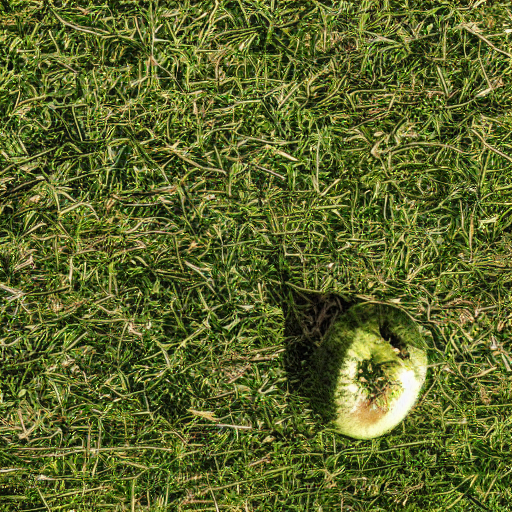}
		\caption{}
	\end{subfigure}
	\begin{subfigure}[b]{0.15\textwidth}
		\includegraphics[width=\textwidth]{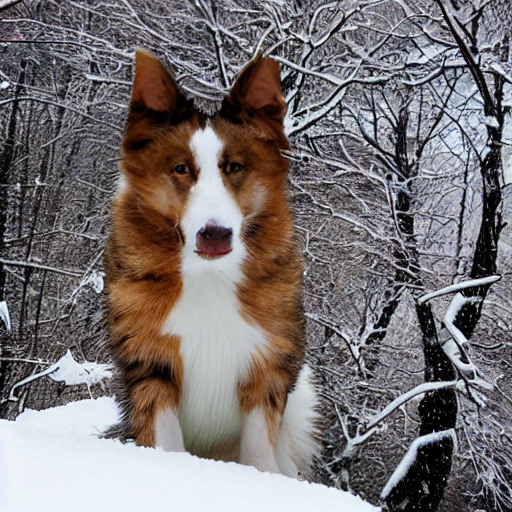}
		\caption{}
	\end{subfigure}
	
	\begin{subfigure}[b]{0.15\textwidth}
		\includegraphics[width=\textwidth]{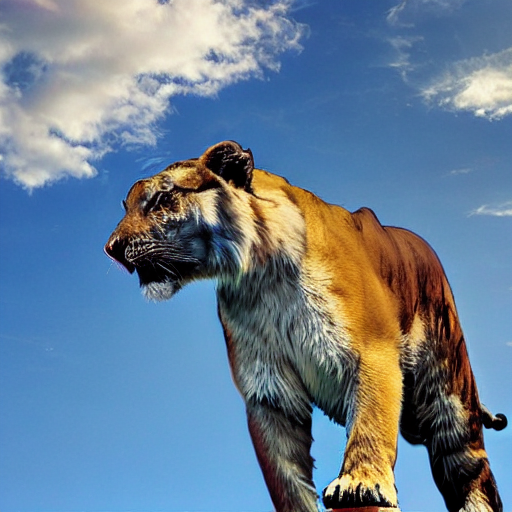}
		\caption{}
	\end{subfigure}
	\begin{subfigure}[b]{0.15\textwidth}
		\includegraphics[width=\textwidth]{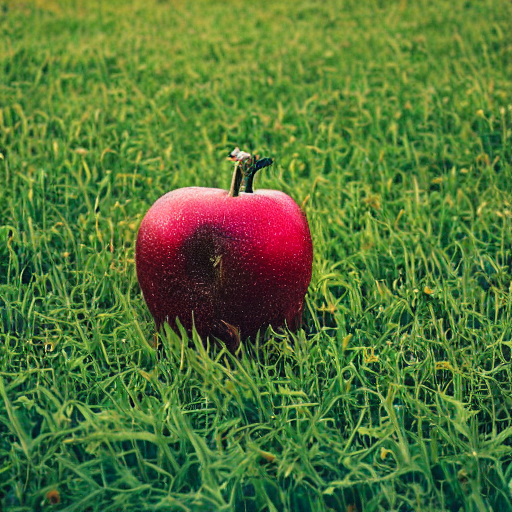}
		\caption{}
	\end{subfigure}
	\begin{subfigure}[b]{0.15\textwidth}
		\includegraphics[width=\textwidth]{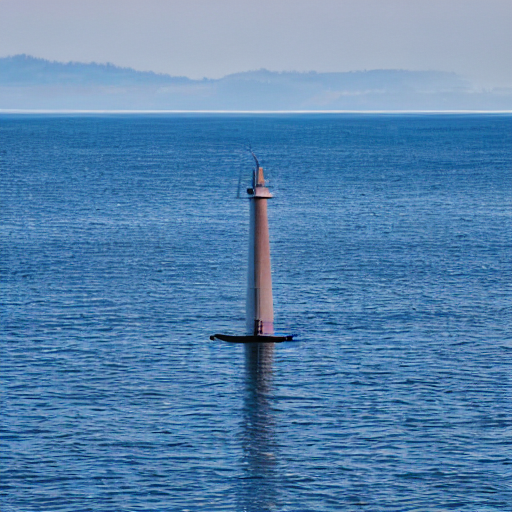}
		\caption{}
	\end{subfigure}
	\begin{subfigure}[b]{0.15\textwidth}
		\includegraphics[width=\textwidth]{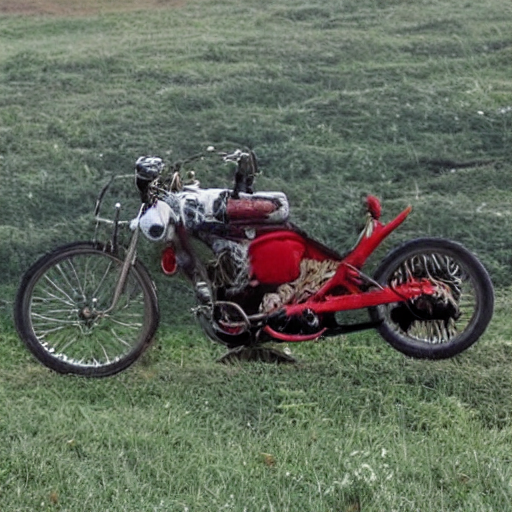}
		\caption{}
	\end{subfigure}
	\begin{subfigure}[b]{0.15\textwidth}
		\includegraphics[width=\textwidth]{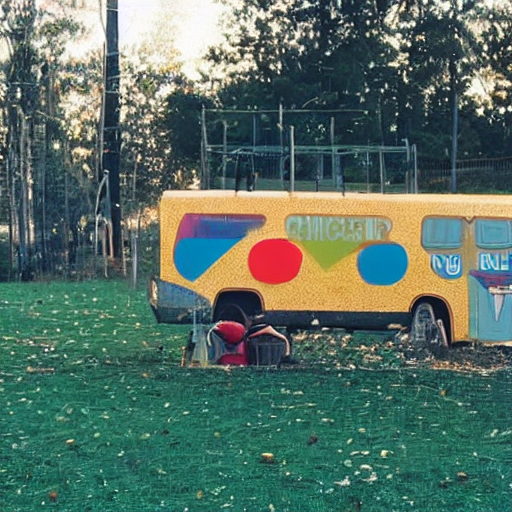}
		\caption{}
	\end{subfigure}
	\caption{Container images}
	\label{fig:cont_img}
\end{figure}

\begin{figure}[htb]
	\centering
	\includegraphics[width=0.6\linewidth]{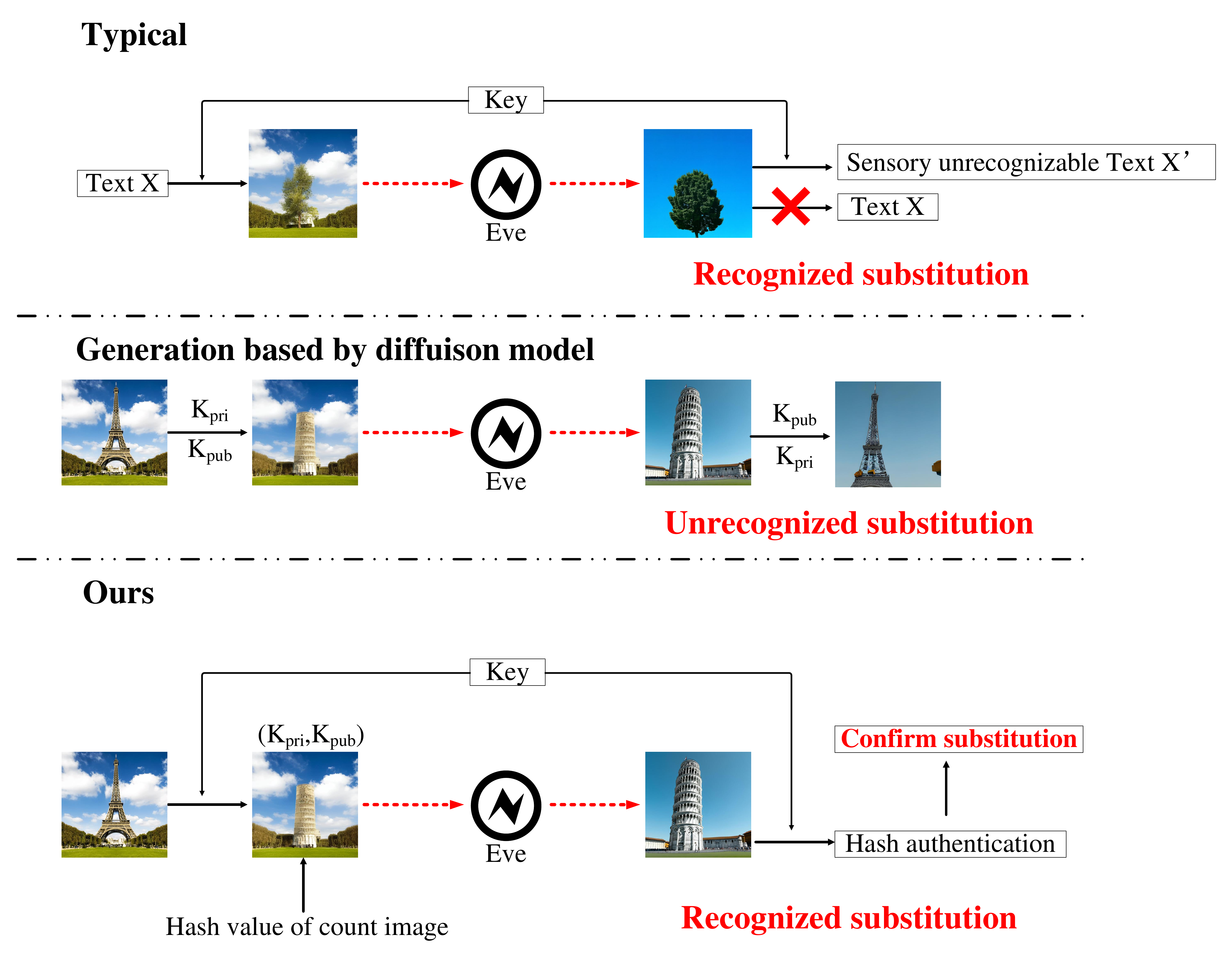}
	\caption{Comparison of steganographic methods in resisting substitution attacks}
	\label{fig:comp_sub_attk}
\end{figure}

\subsection{Resistance to Substitution Attacks}
\textbf{Definition} A substitution attack occurs when an adversary, without access to the communication keys, replaces the stego-object \(x\) sent by the sender with another object \(x'\) of the same format.
If the receiver, using the previously agreed-upon algorithm and keys, fails to detect the alteration and accepts \(x'\) as valid, the attack is considered successful.

In traditional steganography, substitution attacks are inherently difficult to execute because any modification to \(x\) would destroy the embedded information, making it impossible for the receiver to extract meaningful data.
However, in coverless steganography based on diffusion models, substitution attacks become significantly easier to perform.
Since \(x\) and \(x'\) share the same structural format and contain visually plausible content, an attacker can simply replace the transmitted container image, thereby altering the extracted information without directly tampering with the embedding mechanism.

To counter this threat, the proposed method leverages CDJB-PEHS to integrate a cryptographic integrity check alongside the embedded condition information.
Fig.~\ref{fig:comp_sub_attk} provides a comparative analysis of different steganographic schemes in their resistance to substitution attacks. 
While traditional methods either prevent extraction or fail to recognize replacement, our approach successfully identifies tampered images through hash authentication.
Our approach embeds a hash of the container image within the image itself, allowing the receiver to verify whether the stego-object has been altered.
If substitution attack occurs, the extracted condition information becomes sensory-unrecognizable, effectively preventing adversarial exploitation.

\begin{figure}[h]
	\centering
	\includegraphics[width=0.7\linewidth]{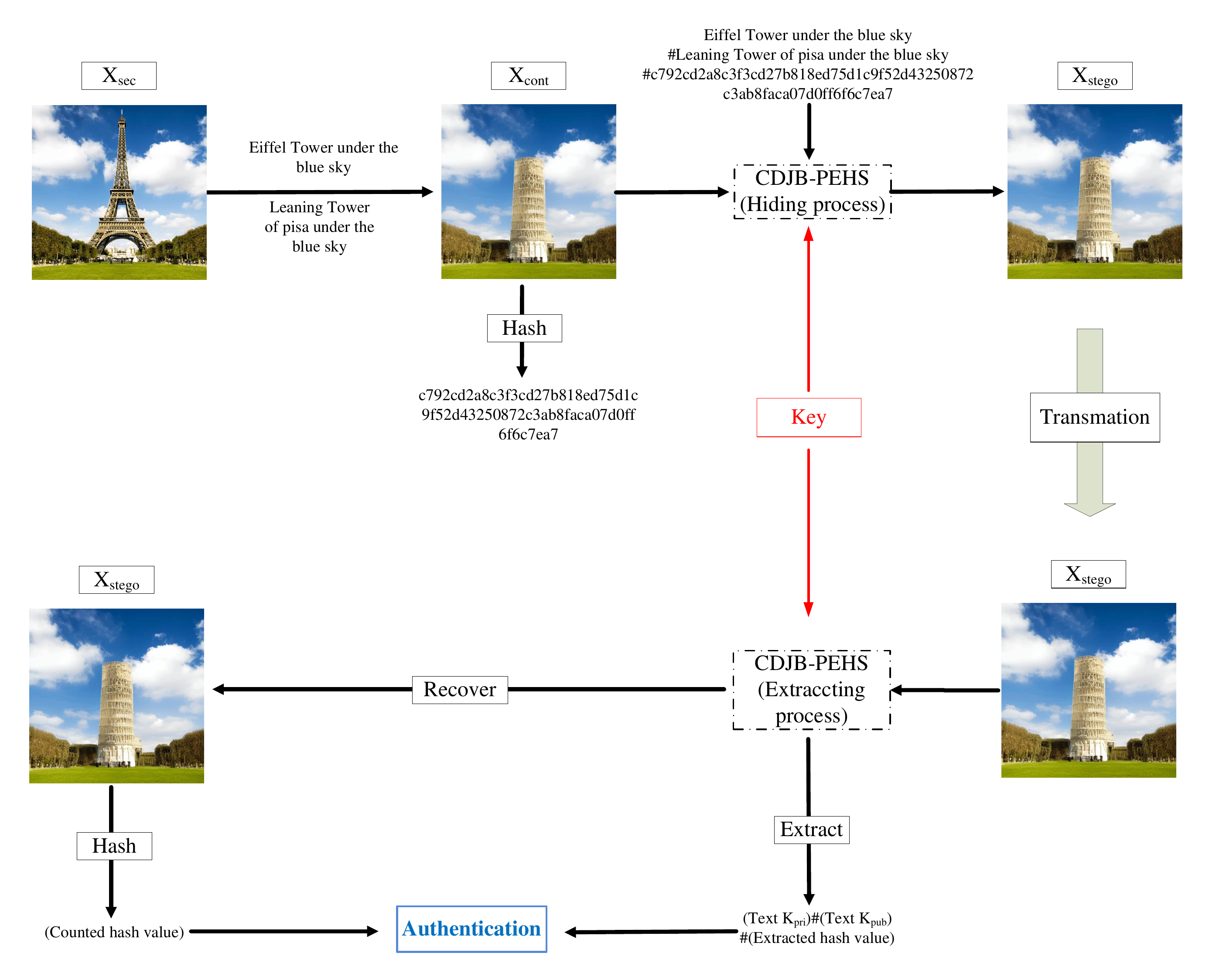}
	\caption{Active defense against substitution attacks}
	\label{fig:def_sub_attk}
\end{figure}

The integrity check is implemented using the SM3 hash function \cite{sm3hash}, which generates a 256-bit hash value along with an 8-bit separator.
According to the analysis in Sec.~\ref{sec:emb_and_sec}, this additional information requires 1320 bits of embedding capacity, which is well within the available space of container images.
The auxiliary data structure follows the format:
\begin{equation}
	K_{pri} \# K_{pub} \# \text{Hash}(X_{cont})
\end{equation}
where each component is separated by \(\#\).
The entire process of appending auxiliary information and performing verification is depicted in Fig.~\ref{fig:def_sub_attk}.

\begin{table}[h]
	\caption{Integrity certification results}
	\centering
	\renewcommand{\arraystretch}{1.2}
	\setlength{\tabcolsep}{5pt}
	
	\begin{tabular}{l c c c c c c c c c c}
		\toprule
		& (a) & (b) & (c) & (d) & (e) & (f) & (g) & (h) & (i) & (j) \\
		\midrule
		\textbf{Correct Stego-object} & \checkmark & \checkmark & \checkmark & \checkmark & \checkmark & \checkmark & \checkmark & \checkmark & \checkmark & \checkmark \\
		\midrule
		\textbf{Substitution Attacked} & \ding{55} & \ding{55} & \ding{55} & \ding{55} & \ding{55} & \ding{55} & \ding{55} & \ding{55} & \ding{55} & \ding{55} \\
		\bottomrule
	\end{tabular}
	\label{tab:check}
\end{table}

\begin{figure}[h]
	\centering
	\includegraphics[width=0.6\linewidth]{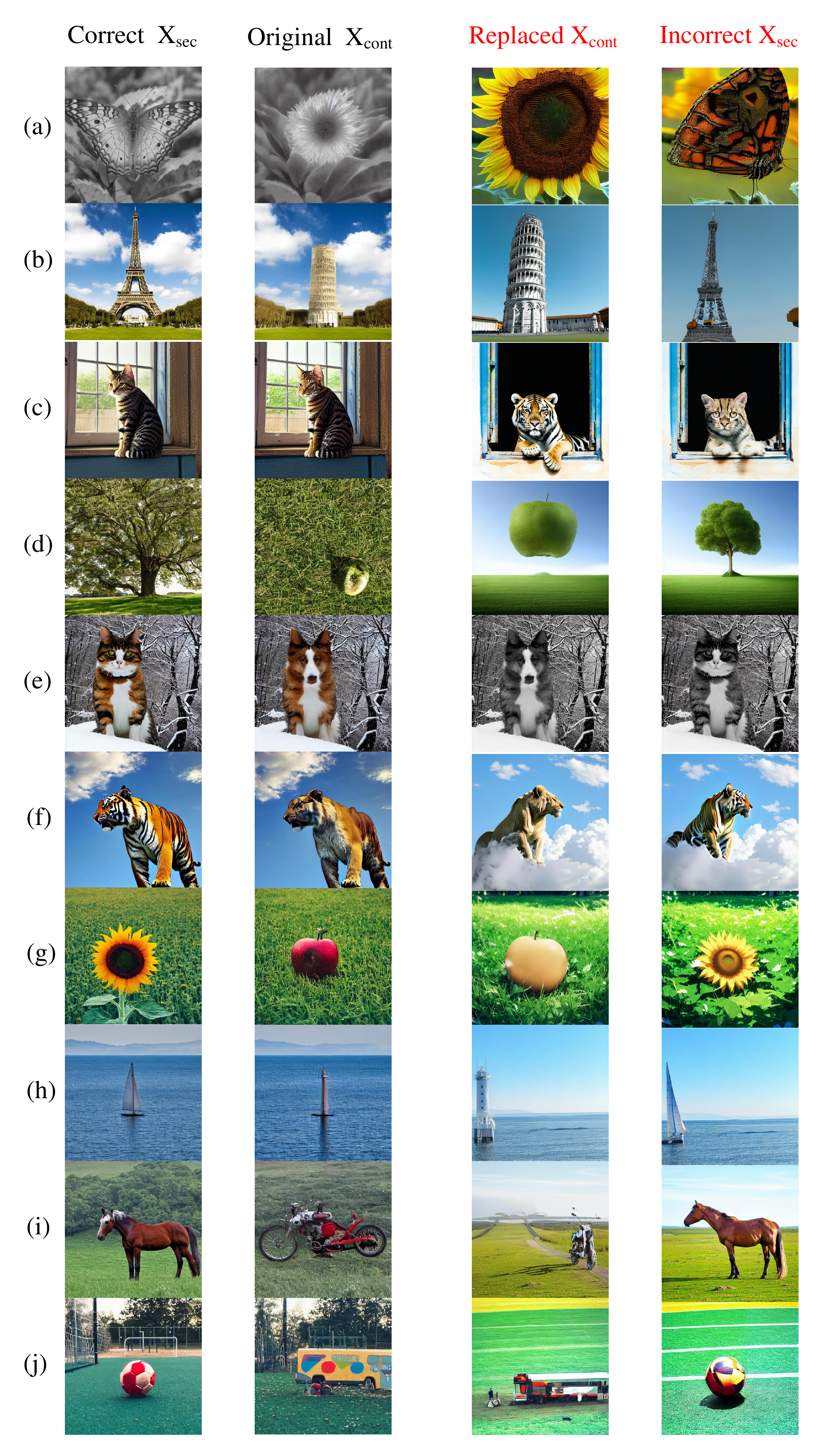}
	\caption{Results of substitution attack}
	\label{fig:res_sub_attk}
\end{figure}

To evaluate the effectiveness of this defense mechanism, we simulate substitution attacks using container images in Fig.~\ref{fig:cont_img}.
Specifically, we consider an attack scenario where an adversary replaces stego-objects generated under the CRoSS \cite{yu2024cross} framework with manipulated versions, using pseudo-keys listed in Table~\ref{tab:similarity}.
The impact of substitution is demonstrated in Fig.~\ref{fig:res_sub_attk}, where incorrect extractions occur in the absence of integrity verification.
Table~\ref{tab:check} shows the verification results for correct stego-objects from Fig.~\ref{fig:pkeys_in_typical} and the incorrect from Fig.~\ref{fig:res_sub_attk}
The "\checkmark" indicates successful verification, while "\ding{55}" indicates failure.
This high success rate is primarily attributed to the properties of the cryptographic hash function and the reversibility of the embedding process.

The avalanche effect of the SM3 hash function ensures that even minimal modifications to an image result in significantly different hash values.
This characteristic makes substitution easily detectable, as an adversary attempting to replace a stego-object with a manipulated version would inevitably introduce inconsistencies in the extracted hash.
Furthermore, since CDJB-PEHS is a reversible steganography method, the original image can be fully reconstructed before extracting the embedded information.
If the stego-object remains unaltered, the extracted image matches the original, leading to a successful hash comparison and confirming the integrity of the transmission.

Despite its effectiveness, the integrity verification mechanism has certain limitations.
While it provides a means of detecting unauthorized modifications, it does not differentiate between intentional attacks and unintended alterations caused by transmission errors.
This makes it particularly suitable for lossless transmission channels, where the primary concern is the prevention of malicious tampering rather than accidental degradation.

In applications such as secure evidence transmission, where images must be protected against any unauthorized modifications, the proposed approach offers a reliable method for ensuring authenticity.
By embedding a cryptographic hash alongside the information, the system ensures that any attempt to manipulate the transmitted image is promptly identified, preserving the integrity and trustworthiness of visual data in high-security scenarios.

\section{Conclusion}
In this paper, we have introduced DDIM-driven coverless steganography scheme that replaces traditional pseudo-keys with a real-key mechanism, significantly enhancing both security and key transmission efficiency.
By integrating Reversible Data Hiding (RDH-PEHS) with chaotic encryption, our approach eliminates the need for frequent key negotiations while reducing the correlation between the key and the secret image.
These improvements strengthen the security of covert communication and make our scheme more practical compared to existing methods.

Beyond its theoretical advantages, this approach offers practical benefits in real-world scenarios where secure image transmission is critical.
For identity protection, it enables anonymization by mapping sensitive facial images to non-identifiable natural scenes.
In confidential data protection, it prevents unauthorized disclosure of restricted visual content, ensuring that sensitive images can be stored and transmitted securely.
Additionally, in legal contexts, the method safeguards evidentiary images against tampering, preserving integrity in forensic and judicial applications.

However, one limitation remains: the scheme’s reliance on the integrity of the stego-object.
If the transmitted image is damaged, the embedded pseudo-key is lost, preventing the recovery of secret information.
While this characteristic inherently enhances integrity verification, future work should focus on improving robustness, potentially by developing alternative verification mechanisms that allow for partial recovery even in the presence of minor modifications.
Further optimizations in embedding techniques could also enhance security while maintaining imperceptibility.
Moreover, exploring broader applications—such as secure identity protection, confidential data transmission, and anti-forensic evidence preservation—could demonstrate the practical value of this approach in high-security environments.

\bibliographystyle{unsrtnat}
\bibliography{references}

\end{document}